\documentclass[a4paper,UKenglish,autoref]{lipics-v2021}

\hideLIPIcs

\addto\extrasUKenglish{

}

\usepackage[dvipsnames]{xcolor}

\usepackage{amsfonts}
\usepackage{mathtools}

\usepackage{subcaption}

\usepackage{threeparttable,booktabs}
\usepackage{tabularx}
\usepackage{makecell}

\usepackage[binary-units=true]{siunitx}
\DeclareSIUnit\persecond{/s}
\newcommand{\Oh}[1]{\mathcal{O}({#1})}
\newcommand{\oh}[1]{o({#1})}
\newcommand{\Th}[1]{\Theta({#1})}
\newcommand{\Wh}[1]{\Omega({#1})}

\newcommand{\ceil}[1]{\lceil #1 \rceil}

\newcommand{\Patrascu}{P\v{a}tra\c{s}cu}

\usepackage{tikz,pgfplots}
\usepgfplotslibrary{groupplots}

\pgfplotscreateplotcyclelist{mycolors}{
    Red, every mark/.append style={solid,scale=0.8}, mark=x \\
    NavyBlue, every mark/.append style={solid,scale=0.9}, mark=+ \\
    ForestGreen, every mark/.append style={solid,scale=0.8}, mark=o \\
    Cyan, every mark/.append style={solid,scale=0.8}, mark=triangle \\
    Plum, every mark/.append style={solid,scale=0.9}, mark=square \\
    BurntOrange, every mark/.append style={solid,scale=0.9}, mark=diamond \\
    RoyalPurple, every mark/.append style={solid,scale=0.8}, mark=x \\
    Sepia, every mark/.append style={solid,scale=0.9}, mark=+ \\
    JungleGreen, every mark/.append style={solid,scale=0.8}, mark=o \\
    Gray, every mark/.append style={solid,scale=0.8}, mark=triangle \\
    RubineRed, every mark/.append style={solid,scale=0.9}, mark=square \\
    Brown, every mark/.append style={solid,scale=0.9}, mark=diamond \\
    Dandelion, every mark/.append style={solid,scale=0.9}, mark=x \\
    Green, every mark/.append style={solid,scale=0.9}, mark=+ \\
    Black, every mark/.append style={solid,scale=0.9}, mark=o \\
}

\pgfplotsset{
    compat=newest,
    cycle list name={mycolors},
    major grid style={thin,dotted=black!50},
    minor grid style={thin,dotted,color=black!50},
    grid,
    xlabel near ticks,
    ylabel near ticks,
    every tick label/.append style={font=\scriptsize},
    every axis label/.style={font=\scriptsize},
    y tick label style={
        xshift=0.5ex,
    },
    x label style={
        yshift=0.5ex,
    },
    y label style={
        yshift=-0.5ex,
    },
    title style={
        font=\scriptsize\bfseries,
        yshift=-1.5ex,
    },
    legend cell align={left},
    legend style={
        inner sep=0.125ex,
        outer sep=0,
        column sep=0,
        font=\scriptsize,
        anchor=north,
        /tikz/every even column/.append style={column sep=3mm,black},
        /tikz/every odd column/.append style={black},
    },
    plot4col/.style={
        width=0.325\textwidth,
        height=0.35\textwidth,
        legend columns=3,
    },
    group3/.style={
        ylabel style={
            align=center,
        },
        title style={
            font=\scriptsize\bfseries,
        },
        width=0.325\textwidth,
        height=0.35\textwidth,
        group style={
            ylabels at=edge left,
            yticklabels at=edge left,
            group size=3 by 3,
            horizontal sep=0em,
        },
        legend columns=3,
    },
    group3wide/.style={
        group3,
        width=0.42\textwidth,
        height=0.375\textwidth,
    },
    lesshigh/.style={
        height=0.2775\textwidth,
    },
    xlognum/.style={
        xtick={27,28,29,30,31,32,33},
        xlabel={\# Keys [$\log_2$]},
    },
    xavglogtput/.style={
        xlabel={Avg. Throughput [$\log_2$ ops/s]},
    },
    ylogtput/.style={
        ylabel={Throughput [$\log_2$ ops/s]},
        ylabel style={align=center},
    },
    ylogmem/.style={
        ylabel={Memory [$\log_2$ B]},
        ylabel style={align=center},
    },
    ybitsperkey/.style={
        ylabel={Memory [bits per key]},
        ylabel style={align=center},
    },
}

\usetikzlibrary{arrows}
\usetikzlibrary{fit}
\usetikzlibrary{matrix}
\usetikzlibrary{positioning}

\newcommand{\qpred}{\text{pred}}

\newcommand{\qselect}{\text{select}}

\newcommand{\bitzero}{\texttt{0}}
\newcommand{\bitone}{\texttt{1}}
\newcommand{\bitdontcare}{\texttt{?}}

\newcommand{\wbranch}{\textsc{Branch}}
\newcommand{\wfree}{\textsc{Free}}
\newcommand{\qmatch}{\text{match}}

\newcommand{\bitaccess}[1]{\ensuremath{\langle #1 \rangle}}
\newcommand{\twodots}{\,..\,}

\newcommand{\msb}{\text{msb}}
\newcommand{\popcnt}{\text{POPCNT}}
\newcommand{\pext}{\text{PEXT}}
\newcommand{\pcmpgtb}{\text{PCMPGTB}}
\newcommand{\tzcnt}{\text{TZCNT}}
\newcommand{\lzcnt}{\text{LZCNT}}

\usepackage{microtype}

\title{Engineering Predecessor Data Structures for Dynamic Integer Sets}

\titlerunning{Engineering Predecessor Data Structures for Dynamic Integer Sets} 

\author{Patrick Dinklage}{TU Dortmund University, Germany}{patrick.dinklage@tu-dortmund.de}{https://orcid.org/0000-0002-2004-6781}{Supported by the German Research Foundation (DFG), priority programme ``Algorithms for Big Data'' (SPP 1736).}
\author{Johannes Fischer}{TU Dortmund University, Germany}{johannes.fischer@cs.tu-dortmund.de}{}{}
\author{Alexander Herlez}{TU Dortmund University, Germany}{alexander.herlez@tu-dortmund.de}{}{Supported by the German Research Foundation (DFG), priority programme ``Algorithms for Big Data'' (SPP 1736).}

\authorrunning{P. Dinklage and J. Fischer and A. Herlez} 

\Copyright{Patrick Dinklage and Johannes Fischer and Alexander Herlez} 



\supplement{\url{https://github.com/pdinklag/tdc/tree/sea21-predecessor}}

\nolinenumbers 

\EventEditors{David Coudert and Emanuele Natale}
\EventNoEds{2}
\EventLongTitle{19th International Symposium on Experimental Algorithms (SEA 2021)}
\EventShortTitle{SEA 2021}
\EventAcronym{SEA}
\EventYear{2021}
\EventDate{June 7--9, 2021}
\EventLocation{Nice, France}
\EventLogo{}
\SeriesVolume{190}
\ArticleNo{7}

\begin{document}

\maketitle

\begin{abstract}
We present highly optimized data structures for the dynamic predecessor problem, where the task is to maintain a set $S$ of $w$-bit numbers under insertions, deletions, and predecessor queries (return the largest element in $S$ no larger than a given key).
The problem of finding predecessors can be viewed as a generalized form of the membership problem, or as a simple version of the nearest neighbour problem. It lies at the core of various real-world problems such as internet routing.

In this work, we engineer (1) a simple implementation of the idea of universe reduction, similar to van-Emde-Boas trees (2) variants of y-fast tries [Willard,~IPL'83], and (3) B-trees with different strategies for organizing the keys contained in the nodes, including an implementation of dynamic fusion nodes [P\v{a}tra\c{s}cu~and~Thorup,~FOCS'14].
We implement our data structures for $w=32,40,64$, which covers most typical scenarios.

Our data structures finish workloads faster than previous approaches while being significantly more space-efficient, e.g., they clearly outperform standard implementations of the STL by finishing up to four times as fast using less than a third of the memory.
Our tests also provide more general insights on data structure design, such as how small sets should be stored and handled and if and when new CPU instructions such as advanced vector extensions pay off.
\end{abstract}

\section{Introduction}
Finding the predecessor of an integer key in a set of keys drawn from a fixed universe is a fundamental algorithmic problem in computer science at the core of real-world applications such as internet routing \cite{DBLP:conf/sigcomm/DegermarkBCP97}.
It can be considered a generalized form of the membership problem or a simple version of the nearest neighbour problem.
Navarro and Rojas-Ledesma \cite{NavarroR20} recently gave a thorough survey on the topic, recapping the past four decades of research.

Data structures for the predecessor problem are designed to beat the $\Wh{\lg n}$ lower time bound for comparison-based searching.
While optimal data structures have been shown for static sets \cite{DBLP:conf/stoc/PatrascuT06} that are known in advance and do not change, they do not necessarily translate to the most practical implementations.
Dinklage et al. \cite{DBLP:conf/esa/Dinklage0HKK20} face the symmetrical \emph{successor} problem for a small universe and develop a simple data structure that accelerates binary search.
Despite not optimal in theory, it is the most efficient in their setting.

In this work, we focus on the \emph{dynamic} problem, where the set of integers can be changed at any time by inserting, deleting or updating keys.
A prominent example of a dynamic predecessor data structure is the \emph{van Emde Boas tree} \cite{DBLP:conf/focs/Boas75}, which, despite near-optimal query times in theory, has been proven irrelevant in practice due to its memory consumption \cite{ma-wenzel}.
Dementiev et al. \cite{DBLP:conf/alenex/DementievKMS04} implemented a \emph{stratified trie} as a heavily simplified practical variant of the van Emde Boas tree for keys drawn from a 32-bit universe.
Nowadays, with 64-bit architectures dominating the landscape, the limitation to 32-bit keys can be considered significant.
The authors gave no hints as to how the data structure can be altered to properly handle larger universes, and simply applying the same structure on a larger universe exceeds practical memory limitations quite quickly.
Nash and Gregg \cite{DBLP:journals/jea/NashG10} thoroughly evaluated various dynamic predecessor data structures in practice, including the aforementioned stratified tree.
They also implemented AVL trees, red-black trees and B-trees, as well as the trie hybrid by Korda and Raman \cite{DBLP:conf/wae/KordaR99} and their self-engineered adaptation of \emph{burst tries} \cite{DBLP:journals/tois/HeinzZW02} for integer keys, which outperform the other data structures regarding both speed and memory usage.

\subparagraph*{Our contributions.}
We engineer new practical solutions for the dynamic predecessor problem that are both faster and more memory efficient than the current best known to us.
First, we apply the idea of \emph{universe sampling} following \cite{DBLP:conf/esa/Dinklage0HKK20} to the dynamic case.
Second, we engineer \emph{y-fast tries} \cite{DBLP:journals/ipl/Willard83}, which, in our view, offer room for many practical optimizations.
Finally, we implement \emph{dynamic fusion nodes} \cite{DBLP:conf/focs/PatrascuT14}, for which \Patrascu{} and Thorup give a very practical description but no implementation.
We embed them into B-trees and make use of modern CPU instructions to accelerate some key low-level operations.

We note that our data structures are designed in a way often not optimal in theory.
A recurring observation that we made is that thanks to large CPU caches, na\"ive solutions for queries on small datasets often outperform sophisticated data structures on modern hardware, including linear scans of unsorted lists, or binary search in na\"ively organized sorted lists, where updates potentially require all items to be shifted.
This observation has been confirmed in the contexts of balanced parentheses \cite{DBLP:conf/wea/BaumstarkGHL17,DBLP:journals/jda/FerradaN17} and finding longest common extensions in strings \cite{DBLP:conf/esa/Dinklage0HKK20,DBLP:conf/spire/IlieT09}.
We make use of this and replace predecessor data structures for small input sets by sorted or unsorted lists without any auxiliary information.

This paper is organized as follows:
we begin with definitions and notations in \autoref{sec:prelim} and a description of our experimental setup in \autoref{sec:methodology}. Then, in Sections~\ref{sec:sampling}--\ref{sec:fusion}, we describe our engineered data structures and give individual experimental results.
In \autoref{sec:comparison}, we conclude with a comparison of the best configurations with existing implementations.

\section{Preliminaries}
\label{sec:prelim}
Let $S$ be a set of $n = |S|$ positive integers called \emph{keys} drawn from a fixed universe $U := [u] = [0, u-1]$.
For any $x \in U$, we call $\qpred_S(x) = \max\{y \in S ~\vert~ y \leq x\}$ the \emph{predecessor} of $x$, which is the largest key in $S$ no larger than $x$.
We consider the dynamic scenario, where keys may be inserted into or deleted from $S$ and the data structure must be updated accordingly.

In our analysis, we use the word RAM model, where we assume that we can perform arithmetic operations on words of size $w = \Theta(\lg u)$ in time $\Oh{1}$ (by default, logarithms are to the base of two).
Additionally, the binary logic operations OR ($\lor$), AND ($\land$) and XOR ($\oplus$) on words take constant time.
With this, we can access the $i$-th bit in a word $x$, denoted by $x\bitaccess{i}$, as well as the bits $i$ to $j$ (inclusively) of $x$, denoted by $x\bitaccess{i \twodots{} j}$, in constant time.
We refer to bit positions in MSBF order, e.g., $x\bitaccess{0}$ is the most significant bit of $x$.

We also require some advanced operations on words to be answered in constant time.
Let $\msb(x)$ denote the position of the most significant set bit of $x$ and $\qselect_1(x, k)$ the position of the $k$-th set bit in $x$.
Another needed operation is counting the number of trailing zero bits of $x$.
In theory, these queries can be answered in constant time using the folklore approach of precomputing universal tables of size $\oh{u^{1/c}}$ bits, where we can look up the answers for all possible queries on a constant number of $c > 1$ blocks of size $w/c$.
In practice, we can make use of special CPU instructions: \lzcnt{} and \tzcnt{} count the number of leading or trailing zeroes, respectively, and \popcnt{} reports the number of set bits in a word.
These instructions are fairly widely spread, being implemented by current versions of the x86-64 (both Intel \cite{intel-instruction-set-reference} and AMD \cite{amd-instruction-set-reference}) instruction sets as well as ARM \cite{arm-instruction-set-reference}.

\subparagraph*{Tries.}

Tries are a long-known information retrieval data structure \cite{Fredkin1960TrieM}.
Here, we consider \emph{binary} tries for strings over a binary alphabet.
Consider a key $x \in U$ to be inserted into a binary trie.
We navigate the trie top-down according to the bits of the binary representation of $x$ in MSBF order:
when reading a 0-bit, we go to the current node's left child, and otherwise to the right child.
Inserting a new element $x$ works accordingly, creating any node that does not yet exist.
Since the binary trie for a set of keys drawn from $U$ has height $\ceil{\lg u}$, this takes total time $\Oh{\lg u}$.
An example of a binary trie is shown in \autoref{fig:trie}.
Binary tries are suitable for solving the dynamic predecessor problem:
to find $\qpred_S(x)$, we navigate down the trie as if we were to insert it.
If we reach a leaf labeled $x$, then $x \in S$ and it is its own predecessor.
Otherwise, we eventually reach an inner node $v$ that is missing the left or right edge that we want to navigate, respectively.
If the right edge is missing, the predecessor of $x$ is the label of the rightmost leaf in the left subtrie of $v$.
Otherwise, if the left edge is missing, we first navigate back up to the lowest ancestor $v'$ of $v$ that has two children and where $v$ is in the right subtrie;
then the predecessor is the label of the rightmost leaf in the left subtrie of $v'$.
In either case, we can report the predecessor of $x$ in time $\Oh{\lg u}$.
Deleting is done by locating a key's leaf, removing it, and navigating back up removing any inner node no longer connected to any leaves, all in time $\Oh{\lg u}$.
The number $\Oh{n\lg u}$ of nodes in the binary trie can be reduced to $\Oh{n}$ by contracting paths of branchless inner nodes to single edges \cite{DBLP:journals/cacm/Maly76}.
We call a trie \emph{compact} if it does not contain any inner nodes with one child.

\section{Methodology}
\label{sec:methodology}

We conduct the following three-step experiment for our data structures:
\begin{enumerate}[(1)]
\item insert $n$ keys drawn uniformly at random from $U$ into the initially empty data structure,
\item perform \emph{ten million} random predecessor queries for keys in the range of the inserted keys, guaranteeing that there is always a predecessor that is never trivially the maximum, and
\item delete the $n$ keys from the data structure in the order in which they were inserted.
\end{enumerate}
In preliminary experiments, we also considered distributions other than uniform, and also intermingling insertions, queries and deletions.
Apart from statistical fluctuations, the results led to the same assertions and thus we solely consider the experiment described above.



For each data structure, we run five iterations using a different random seed each (but the same seeds for all data structures and in the same order).
We measure running times by the wall clock time difference between start and finish of an iteration, as well as the RAM usage using custom overridden versions of \texttt{malloc} and \texttt{free} and compute the averages over the five iterations.
Our code is written in C++17 and publicly available\footnote{Our code is published at \url{https://github.com/pdinklag/tdc/tree/sea21-predecessor}. Make sure to check out the \emph{sea21-predecessor} branch, which contains instructions in the readme.}; we compile using GCC version 9.3.
For hash tables (Sections \ref{sec:sampling} and \ref{sec:yfast}), we use a public\footnote{Robin Hood hashing by Martin Ankerl: \url{https://github.com/martinus/robin-hood-hashing}.} implementation of Robin Hood hashing \cite{DBLP:conf/focs/CelisLM85} that is both faster and more memory efficient than the STL implementation (\texttt{std::unordered\_map}).
We conduct our experiments on Linux machines with an Intel~Xeon~E5-4640v4 processor (12 cores at 2.1~GHz, 12$\times$\SI{32}{\kilo\byte} L1, 12$\times$\SI{256}{\kilo\byte} L2, \SI{30}{\mega\byte} L3 shared, line size \SI{64}{\byte}) and \SI{256}{\giga\byte} of RAM.

\section{Dynamic Universe Sampling}
\label{sec:sampling}

A common technique used by predecessor data structures is known as \emph{length reduction} \cite{DBLP:reference/algo/Belazzougui16}, where we partition the universe into buckets of size $b \ll u$ and reduce the predecessor problem to the much smaller sub-universe $[b]$.
For each bucket, we determine a representative, e.g., the minimum contained key, which is entered into a \emph{top level} predecessor data structure.
The buckets are maintained on the \emph{bucket level}.
When $\qpred(x)$ is queried for some $x \in U$, we first solve the predecessor problem on the top level to find the bucket that $x$ belongs into, and then reduce the query to a smaller one answered on the bucket level.
The van Emde Boas tree \cite{DBLP:conf/focs/Boas75} applies this approach recursively.
In this work, we develop a dynamic version the two-level data structure by Dinklage et al. \cite{DBLP:conf/esa/Dinklage0HKK20} that achieved very good practical results for the \emph{static} predecessor problem.

We partition $U$ into buckets of size $b = 2^k$ for some $k > 0$.
Let $i \in [u/b]$, then the $i$-th bucket can only contain keys from the interval $[bi, b(i+1)-1]$.
We call a bucket \emph{active} if it contains at least one key from $S$.
Since $b = 2^k$, the number $i$ of the bucket that a key $x \in U$ belongs into is the number represented by the $\ceil{\lg u} - k$ highest bits of $x$.
Hence, we only store the lowest $k$ bits for each key to reduce space usage in the buckets.


\subparagraph*{Top level.}
The top level maintains the set of active buckets.
Consider a query for $\qpred_S(x)$, then it reports the \emph{rightmost} active bucket $i$ such that $bi \leq x$.
The predecessor of $x$ is then contained in bucket $i$ if $x$ is greater than the bucket's current minimum.
Otherwise, it is the current maximum key contained in the active bucket preceding $i$.
Clearly, the top level requires a dynamic predecessor data structure on the set of active buckets, i.e., keys drawn from the universe $[u/b]$ represented by the keys' high bits.
We explore two basic options.
First, let $i_{\min}$ and $i_{\max}$ be the numbers of the leftmost and rightmost active buckets, respectively.
We store $i_{\max} - i_{\min} = \Oh{u/b}$ pointers in an array such that the $i$-th entry points to the rightmost active bucket $i'$ with $i' \leq i$.
Predecessor queries can trivially be answered in time $\Oh{1}$ using a lookup, but updates may take time $\Oh{u/b}$ in the worst case as we may need to shift pointers and/or update pointers for succeeding non-active buckets.
Furthermore, the array requires up to $\ceil{(u/b) \lg(u/b)}$ bits of space.
Our alternative is a hash table $H$ containing only pointers to active buckets, identified by their numbers.
Let $b'$ be the number of active buckets, then $H$ requires $\Oh{b' \lg(u/b)}$ bits of space.
Updates can be done in $\Oh{1}$ expected time, but since the order of buckets in $H$ is arbitrary, queries may require to perform up to $b'$ lookups:
when a key belongs in bucket $i$, we look up $i$ in $H$; if that bucket is not active, we find no result and look up $i-1$, and so on.
This takes up to $\Oh{b'}$ expected time.

\subparagraph*{Bucket level.}
On the bucket level, we first look at two basic data structures.
We only store the lowest $k=\lg b$ bits of the contained keys, called \emph{truncated keys} in the following, as the high bits are already defined by the bucket number.
Let $S_i \subseteq S$ be the set of truncated keys contained in the $i$-th bucket.
We can store them in a bit vector $B_i \in \{0,1\}^b$ where $B_i[x]=1$ if $x \in S_i$ and $B_i[x]=0$ otherwise.
Updates are then done in constant time by setting or clearing the respective bit in $B_i$.
To compute $\qpred_{S_i}(x)$, we scan $B_i$ linearly in time $\Oh{b/\lg u}$ using word packing.
However, $B_i$ always requires $b$ bits of space.
Let $n'$ be the \emph{current} number of keys contained in a bucket and consider the case where $n' < b/\lg b$.
An alternative is storing an unsorted list of $n'$ keys:
this requires only $n' \lg b < b$ bits of space and retains $\Oh{1}$ time insertions, and predecessor queries and deletions take time $\Oh{n'} = \Oh{b/\lg b} = \Oh{b/\lg u}$.

\begin{figure}[tb]
\centering
\begin{small}
\begin{tikzpicture}[
    top/.style={draw,shape=rectangle,minimum width=8em},
    bucket/.style={draw,shape=rectangle,minimum width=8em,inner sep=0.25em},
]
\node [top] (top-0) at (0, 0) { [0, 7] };
\node [top] (top-1) at (8em, 0) { [8, 15] };
\node [top] (top-2) at (16em, 0) { [16, 23] };
\node [top] (top-3) at (24em, 0) { [24, 31] };

\node [bucket] (bucket-0) at (0, -2.5em) { 01101011 };
\node [bucket] (bucket-2) at (16em, -2.5em) { 5, 3 };

\draw[-latex] (top-0.south) to (bucket-0.north);
\draw[-latex] (top-1.south) to (bucket-0.north);
\draw[-latex] (top-2.south) to (bucket-2.north);
\draw[-latex] (top-3.south) to (bucket-2.north);

\node [right=0 of bucket-0] {$B_1$};
\node [right=0 of bucket-2] {$B_2$};

\node [left=0.5em of top-0] {Top:};
\node [left=0.5em of bucket-0] {Buckets:};

\end{tikzpicture}
\end{small}
\caption{
    Hybrid universe sampling data structure for $S=\{ 1, 2, 4, 6, 7, 19, 21 \}$ with $w=5$, $b=8$ and $\theta_{\min} = \theta_{\max} = 3$ .
    The top level holds bucket pointers for the partitioned universe.
    Since there are no keys from the intervals $[8,15]$ and $[24,31]$ contained in $S$, their pointers point to the respective preceding buckets.
    Bucket $B_1$ is represented as a bit vector of length $b$ such that each 1-bit corresponds to a key contained in $S$.
    Bucket $B_2$, on the other hand, only contains two keys that are represented as an unsorted list of keys relative to the left interval boundary.
}
\label{fig:hybrid-sampling}
\end{figure}
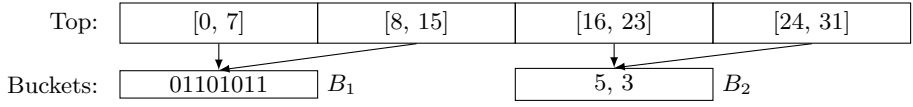

We now consider a hybrid of the two basic bucket structures.
Let $\theta_{\min}$ and $\theta_{\max}$ be thresholds with $0 < \theta_{\min} \leq \theta_{\max} < b$.
We maintain a bucket as an unsorted list of keys as long as $n' < \theta_{\max}$.
If, after inserts, the bucket grows beyond $\theta_{\max}$ keys, we rebuild it to a bit vector.
If, after deletions, the bucket size falls below $\theta_{\min}$ keys, we revert to an unsorted list.
Rebuilding the bucket to a bit vector or unsorted list, respectively, takes time $\Oh{b}$.
Let $\theta_{\min} := c b/\lg b$ and $\theta_{\max} := \theta_{\min} + c' \lg b$ for constants $c, c' > 0$.
Then, $\Th{\lg b}$ insertions need to occur before we switch to a bit vector representation, followed by $\Th{\lg b}$ deletions before reverting to an unsorted list.
We can thus amortize the time needed for one insertion and one deletion to $\Oh{b/\lg b}$.
At all times, predecessor queries take at most $\Oh{b/\lg u}$ time and the bucket requires at most $b$ bits of space.
\autoref{fig:hybrid-sampling} shows an example.

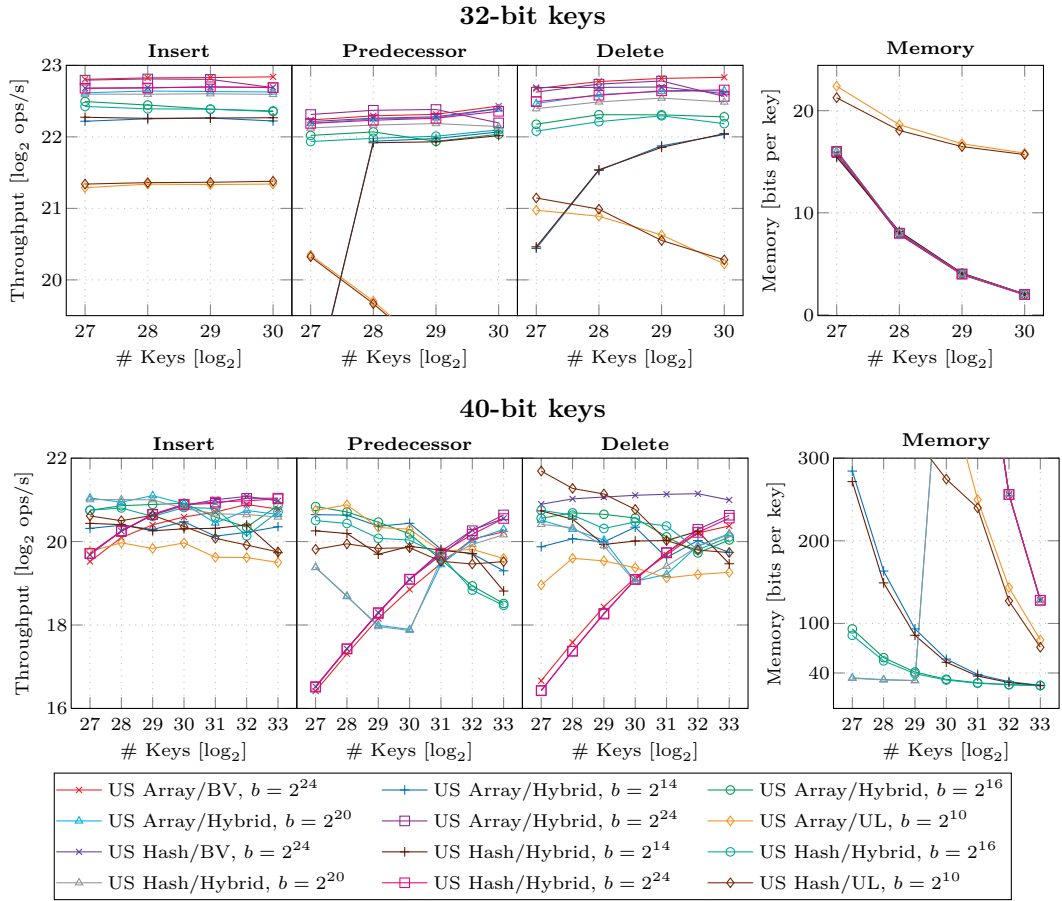
\begin{figure}[tb]
\hspace*{\fill}\textbf{32-bit keys}\hspace*{\fill}

\begin{tikzpicture}
\begin{groupplot}[group3, ylogtput, ymin=19.5, ymax=23]
\nextgroupplot[xlognum, title={Insert}, legend to name=leg:sampling:u40]
\addplot coordinates { (27.0,22.8046) (28.0,22.825) (29.0,22.8294) (30.0,22.8395) };
\addlegendentry{US Array/BV, $b=2^{24}$};
\addplot coordinates { (27.0,22.2183) (28.0,22.2527) (29.0,22.2631) (30.0,22.2225) };
\addlegendentry{US Array/Hybrid, $b=2^{14}$};
\addplot coordinates { (27.0,22.4943) (28.0,22.4445) (29.0,22.3875) (30.0,22.354) };
\addlegendentry{US Array/Hybrid, $b=2^{16}$};
\addplot coordinates { (27.0,22.6156) (28.0,22.6417) (29.0,22.6336) (30.0,22.6279) };
\addlegendentry{US Array/Hybrid, $b=2^{20}$};
\addplot coordinates { (27.0,22.7934) (28.0,22.81) (29.0,22.8037) (30.0,22.6882) };
\addlegendentry{US Array/Hybrid, $b=2^{24}$};
\addplot coordinates { (27.0,21.2896) (28.0,21.3389) (29.0,21.333) (30.0,21.3403) };
\addlegendentry{US Array/UL, $b=2^{10}$};
\addplot coordinates { (27.0,22.6812) (28.0,22.6878) (29.0,22.6955) (30.0,22.6926) };
\addlegendentry{US Hash/BV, $b=2^{24}$};
\addplot coordinates { (27.0,22.2741) (28.0,22.2585) (29.0,22.2633) (30.0,22.2677) };
\addlegendentry{US Hash/Hybrid, $b=2^{14}$};
\addplot coordinates { (27.0,22.4254) (28.0,22.3919) (29.0,22.3871) (30.0,22.3634) };
\addlegendentry{US Hash/Hybrid, $b=2^{16}$};
\addplot coordinates { (27.0,22.5968) (28.0,22.5964) (29.0,22.6029) (30.0,22.5937) };
\addlegendentry{US Hash/Hybrid, $b=2^{20}$};
\addplot coordinates { (27.0,22.6797) (28.0,22.6834) (29.0,22.7011) (30.0,22.6886) };
\addlegendentry{US Hash/Hybrid, $b=2^{24}$};
\addplot coordinates { (27.0,21.3396) (28.0,21.3602) (29.0,21.3649) (30.0,21.3789) };
\addlegendentry{US Hash/UL, $b=2^{10}$};
\nextgroupplot[xlognum, title={Predecessor}]
\addplot coordinates { (27.0,22.2365) (28.0,22.2938) (29.0,22.3154) (30.0,22.4308) };
\addlegendentry{US Array/BV, $b=2^{24}$};
\addplot coordinates { (27.0,18.2507) (28.0,21.9389) (29.0,21.9767) (30.0,22.0689) };
\addlegendentry{US Array/Hybrid, $b=2^{14}$};
\addplot coordinates { (27.0,22.0203) (28.0,22.0691) (29.0,21.9377) (30.0,22.0362) };
\addlegendentry{US Array/Hybrid, $b=2^{16}$};
\addplot coordinates { (27.0,22.189) (28.0,22.2516) (29.0,22.2649) (30.0,22.3914) };
\addlegendentry{US Array/Hybrid, $b=2^{20}$};
\addplot coordinates { (27.0,22.313) (28.0,22.3717) (29.0,22.3829) (30.0,22.1953) };
\addlegendentry{US Array/Hybrid, $b=2^{24}$};
\addplot coordinates { (27.0,20.3475) (28.0,19.7022) (29.0,18.9421) (30.0,18.059) };
\addlegendentry{US Array/UL, $b=2^{10}$};
\addplot coordinates { (27.0,22.214) (28.0,22.2604) (29.0,22.2808) (30.0,22.397) };
\addlegendentry{US Hash/BV, $b=2^{24}$};
\addplot coordinates { (27.0,18.2915) (28.0,21.9178) (29.0,21.9311) (30.0,22.0164) };
\addlegendentry{US Hash/Hybrid, $b=2^{14}$};
\addplot coordinates { (27.0,21.9348) (28.0,21.9797) (29.0,22.009) (30.0,22.0969) };
\addlegendentry{US Hash/Hybrid, $b=2^{16}$};
\addplot coordinates { (27.0,22.1239) (28.0,22.1652) (29.0,22.1863) (30.0,22.1438) };
\addlegendentry{US Hash/Hybrid, $b=2^{20}$};
\addplot coordinates { (27.0,22.1878) (28.0,22.2324) (29.0,22.2602) (30.0,22.3573) };
\addlegendentry{US Hash/Hybrid, $b=2^{24}$};
\addplot coordinates { (27.0,20.3234) (28.0,19.6672) (29.0,18.9191) (30.0,18.0516) };
\addlegendentry{US Hash/UL, $b=2^{10}$};
\legend{}
\nextgroupplot[xlognum, title={Delete}]
\addplot coordinates { (27.0,22.6753) (28.0,22.7734) (29.0,22.8174) (30.0,22.8328) };
\addlegendentry{US Array/BV, $b=2^{24}$};
\addplot coordinates { (27.0,20.4402) (28.0,21.5313) (29.0,21.875) (30.0,22.0328) };
\addlegendentry{US Array/Hybrid, $b=2^{14}$};
\addplot coordinates { (27.0,22.1761) (28.0,22.31) (29.0,22.3079) (30.0,22.2798) };
\addlegendentry{US Array/Hybrid, $b=2^{16}$};
\addplot coordinates { (27.0,22.4662) (28.0,22.5919) (29.0,22.6345) (30.0,22.6452) };
\addlegendentry{US Array/Hybrid, $b=2^{20}$};
\addplot coordinates { (27.0,22.626) (28.0,22.7379) (29.0,22.7815) (30.0,22.5656) };
\addlegendentry{US Array/Hybrid, $b=2^{24}$};
\addplot coordinates { (27.0,20.9744) (28.0,20.89) (29.0,20.6288) (30.0,20.2226) };
\addlegendentry{US Array/UL, $b=2^{10}$};
\addplot coordinates { (27.0,22.6899) (28.0,22.6917) (29.0,22.6973) (30.0,22.6043) };
\addlegendentry{US Hash/BV, $b=2^{24}$};
\addplot coordinates { (27.0,20.462) (28.0,21.5398) (29.0,21.853) (30.0,22.0451) };
\addlegendentry{US Hash/Hybrid, $b=2^{14}$};
\addplot coordinates { (27.0,22.0797) (28.0,22.2126) (29.0,22.2945) (30.0,22.1847) };
\addlegendentry{US Hash/Hybrid, $b=2^{16}$};
\addplot coordinates { (27.0,22.3941) (28.0,22.4886) (29.0,22.5422) (30.0,22.4875) };
\addlegendentry{US Hash/Hybrid, $b=2^{20}$};
\addplot coordinates { (27.0,22.4923) (28.0,22.5822) (29.0,22.6435) (30.0,22.656) };
\addlegendentry{US Hash/Hybrid, $b=2^{24}$};
\addplot coordinates { (27.0,21.1456) (28.0,20.9889) (29.0,20.5512) (30.0,20.2809) };
\addlegendentry{US Hash/UL, $b=2^{10}$};
\legend{}
\end{groupplot}
\end{tikzpicture}
\begin{tikzpicture}
\begin{axis}[plot4col, xlognum, ybitsperkey, title={Memory}]
\addplot coordinates { (27.0,15.7503) (28.0,7.87517) (29.0,3.93759) (30.0,1.96879) };
\addlegendentry{US Array/BV, $b=2^{24}$};
\addplot coordinates { (27.0,15.4985) (28.0,8.18681) (29.0,4.0934) (30.0,2.0467) };
\addlegendentry{US Array/Hybrid, $b=2^{14}$};
\addplot coordinates { (27.0,16.0934) (28.0,8.0467) (29.0,4.02335) (30.0,2.01167) };
\addlegendentry{US Array/Hybrid, $b=2^{16}$};
\addplot coordinates { (27.0,16.0061) (28.0,8.00304) (29.0,4.00152) (30.0,2.00076) };
\addlegendentry{US Array/Hybrid, $b=2^{20}$};
\addplot coordinates { (27.0,16.0004) (28.0,8.00019) (29.0,4.00009) (30.0,2.00005) };
\addlegendentry{US Array/Hybrid, $b=2^{24}$};
\addplot coordinates { (27.0,22.3965) (28.0,18.6289) (29.0,16.765) (30.0,15.8419) };
\addlegendentry{US Array/UL, $b=2^{10}$};
\addplot coordinates { (27.0,15.7505) (28.0,7.87524) (29.0,3.93762) (30.0,1.96881) };
\addlegendentry{US Hash/BV, $b=2^{24}$};
\addplot coordinates { (27.0,15.4467) (28.0,8.16094) (29.0,4.08047) (30.0,2.04024) };
\addlegendentry{US Hash/Hybrid, $b=2^{14}$};
\addplot coordinates { (27.0,16.0794) (28.0,8.03971) (29.0,4.01986) (30.0,2.00993) };
\addlegendentry{US Hash/Hybrid, $b=2^{16}$};
\addplot coordinates { (27.0,16.0051) (28.0,8.00257) (29.0,4.00129) (30.0,2.00064) };
\addlegendentry{US Hash/Hybrid, $b=2^{20}$};
\addplot coordinates { (27.0,16.0005) (28.0,8.00024) (29.0,4.00012) (30.0,2.00006) };
\addlegendentry{US Hash/Hybrid, $b=2^{24}$};
\addplot coordinates { (27.0,21.2665) (28.0,18.0639) (29.0,16.4825) (30.0,15.7007) };
\addlegendentry{US Hash/UL, $b=2^{10}$};
\legend{}
\end{axis}
\end{tikzpicture}

\hspace*{\fill}\textbf{40-bit keys}\hspace*{\fill}

\begin{tikzpicture}
\begin{groupplot}[group3, ylogtput, ymin=16, ymax=22]
\nextgroupplot[xlognum, title={Insert}]
\addplot coordinates { (27.0,19.5223) (28.0,20.0953) (29.0,20.401) (30.0,20.5888) (31.0,20.7065) (32.0,20.889) (33.0,20.779) };
\addlegendentry{US Array/BV, $b=2^{24}$};
\addplot coordinates { (27.0,20.3156) (28.0,20.3919) (29.0,20.2909) (30.0,20.4756) (31.0,20.1293) (32.0,20.2232) (33.0,20.3546) };
\addlegendentry{US Array/Hybrid, $b=2^{14}$};
\addplot coordinates { (27.0,20.7507) (28.0,20.8594) (29.0,20.8898) (30.0,20.9175) (31.0,20.5861) (32.0,20.3463) (33.0,20.8644) };
\addlegendentry{US Array/Hybrid, $b=2^{16}$};
\addplot coordinates { (27.0,21.0457) (28.0,20.9435) (29.0,21.1026) (30.0,20.9061) (31.0,20.4452) (32.0,20.746) (33.0,20.6496) };
\addlegendentry{US Array/Hybrid, $b=2^{20}$};
\addplot coordinates { (27.0,19.718) (28.0,20.2412) (29.0,20.6545) (30.0,20.8835) (31.0,20.9173) (32.0,21.0373) (33.0,21.0404) };
\addlegendentry{US Array/Hybrid, $b=2^{24}$};
\addplot coordinates { (27.0,19.7666) (28.0,19.9704) (29.0,19.835) (30.0,19.9666) (31.0,19.6245) (32.0,19.617) (33.0,19.4991) };
\addlegendentry{US Array/UL, $b=2^{10}$};
\addplot coordinates { (27.0,19.6852) (28.0,20.2519) (29.0,20.6337) (30.0,20.8641) (31.0,21.0027) (32.0,21.0836) (33.0,20.9753) };
\addlegendentry{US Hash/BV, $b=2^{24}$};
\addplot coordinates { (27.0,20.4366) (28.0,20.4061) (29.0,20.2586) (30.0,20.3062) (31.0,20.3178) (32.0,20.3995) (33.0,19.7323) };
\addlegendentry{US Hash/Hybrid, $b=2^{14}$};
\addplot coordinates { (27.0,20.7502) (28.0,20.8064) (29.0,20.6015) (30.0,20.8168) (31.0,20.7945) (32.0,20.1368) (33.0,20.7149) };
\addlegendentry{US Hash/Hybrid, $b=2^{16}$};
\addplot coordinates { (27.0,21.005) (28.0,21.0066) (29.0,21.0015) (30.0,20.7972) (31.0,20.6667) (32.0,20.6482) (33.0,20.5891) };
\addlegendentry{US Hash/Hybrid, $b=2^{20}$};
\addplot coordinates { (27.0,19.7131) (28.0,20.2558) (29.0,20.6466) (30.0,20.8934) (31.0,20.9478) (32.0,20.9777) (33.0,21.0236) };
\addlegendentry{US Hash/Hybrid, $b=2^{24}$};
\addplot coordinates { (27.0,20.6144) (28.0,20.5002) (29.0,20.6304) (30.0,20.3635) (31.0,20.079) (32.0,19.9173) (33.0,19.7513) };
\addlegendentry{US Hash/UL, $b=2^{10}$};
\legend{}
\nextgroupplot[xlognum, title={Predecessor}]
\addplot coordinates { (27.0,16.4165) (28.0,17.3085) (29.0,18.1548) (30.0,18.853) (31.0,19.4745) (32.0,20.0582) (33.0,20.2341) };
\addlegendentry{US Array/BV, $b=2^{24}$};
\addplot coordinates { (27.0,20.6458) (28.0,20.634) (29.0,20.3683) (30.0,20.4384) (31.0,19.8171) (32.0,19.7083) (33.0,19.2991) };
\addlegendentry{US Array/Hybrid, $b=2^{14}$};
\addplot coordinates { (27.0,20.8397) (28.0,20.7096) (29.0,20.4662) (30.0,20.1824) (31.0,19.5675) (32.0,18.9374) (33.0,18.5148) };
\addlegendentry{US Array/Hybrid, $b=2^{16}$};
\addplot coordinates { (27.0,19.3778) (28.0,18.6827) (29.0,17.9944) (30.0,17.8929) (31.0,19.4477) (32.0,20.0356) (33.0,20.293) };
\addlegendentry{US Array/Hybrid, $b=2^{20}$};
\addplot coordinates { (27.0,16.5094) (28.0,17.4242) (29.0,18.2984) (30.0,19.0932) (31.0,19.677) (32.0,20.2616) (33.0,20.6391) };
\addlegendentry{US Array/Hybrid, $b=2^{24}$};
\addplot coordinates { (27.0,20.7477) (28.0,20.8805) (29.0,20.3389) (30.0,20.2964) (31.0,19.7681) (32.0,19.8138) (33.0,19.5942) };
\addlegendentry{US Array/UL, $b=2^{10}$};
\addplot coordinates { (27.0,16.5218) (28.0,17.4281) (29.0,18.296) (30.0,19.0934) (31.0,19.7637) (32.0,20.276) (33.0,20.554) };
\addlegendentry{US Hash/BV, $b=2^{24}$};
\addplot coordinates { (27.0,20.2567) (28.0,20.1934) (29.0,19.6964) (30.0,19.8853) (31.0,19.7948) (32.0,19.7112) (33.0,18.8143) };
\addlegendentry{US Hash/Hybrid, $b=2^{14}$};
\addplot coordinates { (27.0,20.5037) (28.0,20.4339) (29.0,20.0786) (30.0,20.0359) (31.0,19.6629) (32.0,18.8349) (33.0,18.4701) };
\addlegendentry{US Hash/Hybrid, $b=2^{16}$};
\addplot coordinates { (27.0,19.383) (28.0,18.6957) (29.0,17.9623) (30.0,17.8691) (31.0,19.5584) (32.0,19.9302) (33.0,20.1642) };
\addlegendentry{US Hash/Hybrid, $b=2^{20}$};
\addplot coordinates { (27.0,16.5247) (28.0,17.4322) (29.0,18.2751) (30.0,19.0997) (31.0,19.7217) (32.0,20.1727) (33.0,20.5527) };
\addlegendentry{US Hash/Hybrid, $b=2^{24}$};
\addplot coordinates { (27.0,19.8147) (28.0,19.9455) (29.0,19.8375) (30.0,19.8526) (31.0,19.5288) (32.0,19.4621) (33.0,19.5196) };
\addlegendentry{US Hash/UL, $b=2^{10}$};
\legend{}
\nextgroupplot[xlognum, title={Delete}]
\addplot coordinates { (27.0,16.665) (28.0,17.5844) (29.0,18.4364) (30.0,19.1344) (31.0,19.6809) (32.0,20.206) (33.0,20.3747) };
\addlegendentry{US Array/BV, $b=2^{24}$};
\addplot coordinates { (27.0,19.8754) (28.0,20.0672) (29.0,19.9903) (30.0,20.3453) (31.0,19.617) (32.0,20.0237) (33.0,19.741) };
\addlegendentry{US Array/Hybrid, $b=2^{14}$};
\addplot coordinates { (27.0,20.5593) (28.0,20.6875) (29.0,20.6535) (30.0,20.5639) (31.0,20.1138) (32.0,19.7273) (33.0,20.0384) };
\addlegendentry{US Array/Hybrid, $b=2^{16}$};
\addplot coordinates { (27.0,20.5125) (28.0,20.2949) (29.0,20.0327) (30.0,19.0473) (31.0,19.2131) (32.0,19.8999) (33.0,20.1886) };
\addlegendentry{US Array/Hybrid, $b=2^{20}$};
\addplot coordinates { (27.0,16.4225) (28.0,17.3728) (29.0,18.2767) (30.0,19.0865) (31.0,19.729) (32.0,20.2922) (33.0,20.631) };
\addlegendentry{US Array/Hybrid, $b=2^{24}$};
\addplot coordinates { (27.0,18.9601) (28.0,19.5995) (29.0,19.536) (30.0,19.3735) (31.0,19.1283) (32.0,19.2119) (33.0,19.2631) };
\addlegendentry{US Array/UL, $b=2^{10}$};
\addplot coordinates { (27.0,20.8944) (28.0,21.0244) (29.0,21.0678) (30.0,21.1092) (31.0,21.1341) (32.0,21.1473) (33.0,20.9974) };
\addlegendentry{US Hash/BV, $b=2^{24}$};
\addplot coordinates { (27.0,20.7357) (28.0,20.5369) (29.0,19.9301) (30.0,20.0136) (31.0,20.0243) (32.0,20.2478) (33.0,19.4686) };
\addlegendentry{US Hash/Hybrid, $b=2^{14}$};
\addplot coordinates { (27.0,20.7506) (28.0,20.6328) (29.0,20.3135) (30.0,20.4709) (31.0,20.3736) (32.0,19.8012) (33.0,20.1097) };
\addlegendentry{US Hash/Hybrid, $b=2^{16}$};
\addplot coordinates { (27.0,20.4129) (28.0,20.3321) (29.0,19.8987) (30.0,19.0254) (31.0,19.4102) (32.0,19.8836) (33.0,20.1691) };
\addlegendentry{US Hash/Hybrid, $b=2^{20}$};
\addplot coordinates { (27.0,16.4305) (28.0,17.374) (29.0,18.2608) (30.0,19.0933) (31.0,19.7381) (32.0,20.2194) (33.0,20.5627) };
\addlegendentry{US Hash/Hybrid, $b=2^{24}$};
\addplot coordinates { (27.0,21.6916) (28.0,21.2743) (29.0,21.1407) (30.0,20.7704) (31.0,20.0244) (32.0,19.7962) (33.0,19.7368) };
\addlegendentry{US Hash/UL, $b=2^{10}$};
\legend{}
\end{groupplot}
\end{tikzpicture}
\begin{tikzpicture}
\begin{axis}[plot4col, xlognum, ybitsperkey, ymax=300, ytick={40,100,200,300}, title={Memory}]
\addplot coordinates { (27.0,8191.93) (28.0,4095.96) (29.0,2047.98) (30.0,1023.99) (31.0,511.996) (32.0,255.998) (33.0,127.999) };
\addlegendentry{US Array/BV, $b=2^{24}$};
\addplot coordinates { (27.0,284.342) (28.0,163.386) (29.0,93.3795) (30.0,56.8302) (31.0,38.2338) (32.0,29.4846) (33.0,25.146) };
\addlegendentry{US Array/Hybrid, $b=2^{14}$};
\addplot coordinates { (27.0,93.3505) (28.0,58.5666) (29.0,41.1369) (30.0,32.4174) (31.0,28.0984) (32.0,25.9697) (33.0,24.9296) };
\addlegendentry{US Array/Hybrid, $b=2^{16}$};
\addplot coordinates { (27.0,34.1305) (28.0,31.9607) (29.0,30.9111) (30.0,539.267) (31.0,512.194) (32.0,256.097) (33.0,128.048) };
\addlegendentry{US Array/Hybrid, $b=2^{20}$};
\addplot coordinates { (27.0,8192.19) (28.0,4096.1) (29.0,2048.05) (30.0,1024.02) (31.0,512.012) (32.0,256.006) (33.0,128.003) };
\addlegendentry{US Array/Hybrid, $b=2^{24}$};
\addplot coordinates { (27.0,1290.93) (28.0,831.321) (29.0,570.08) (30.0,392.229) (31.0,249.793) (32.0,143.422) (33.0,80.099) };
\addlegendentry{US Array/UL, $b=2^{10}$};
\addplot coordinates { (27.0,8191.92) (28.0,4095.96) (29.0,2047.98) (30.0,1023.99) (31.0,511.995) (32.0,255.997) (33.0,127.999) };
\addlegendentry{US Hash/BV, $b=2^{24}$};
\addplot coordinates { (27.0,271.596) (28.0,149.058) (29.0,85.6042) (30.0,52.9369) (31.0,36.2871) (32.0,28.5113) (33.0,24.6593) };
\addlegendentry{US Hash/Hybrid, $b=2^{14}$};
\addplot coordinates { (27.0,85.5791) (28.0,54.6751) (29.0,39.1912) (30.0,31.4445) (31.0,27.612) (32.0,25.7265) (33.0,24.8079) };
\addlegendentry{US Hash/Hybrid, $b=2^{16}$};
\addplot coordinates { (27.0,33.5253) (28.0,31.6581) (29.0,30.7598) (30.0,539.191) (31.0,512.156) (32.0,256.078) (33.0,128.039) };
\addlegendentry{US Hash/Hybrid, $b=2^{20}$};
\addplot coordinates { (27.0,8192.17) (28.0,4096.08) (29.0,2048.04) (30.0,1024.02) (31.0,512.01) (32.0,256.005) (33.0,128.003) };
\addlegendentry{US Hash/Hybrid, $b=2^{24}$};
\addplot coordinates { (27.0,480.985) (28.0,477.452) (29.0,323.897) (30.0,274.575) (31.0,239.766) (32.0,127.187) (33.0,71.1248) };
\addlegendentry{US Hash/UL, $b=2^{10}$};
\legend{}
\end{axis}
\end{tikzpicture}
\centering\ref{leg:sampling:u40}
\caption{
Throughputs for the insert, predecessor and delete operations, as well as memory usage of the universe sampling data structures for 32-bit (top) and 40-bit keys (bottom).
Best viewed in colour.
In the legend, \emph{US} stands for universe sampling, \emph{BV} stands for buckets backed by bit vectors, \emph{UL} for unsorted lists.
Missing points indicate throughputs lower than $2^{20}$ operations per second or exceeding of 300 bits consumed per key, respectively, and are omitted for clarity.
}
\label{fig:sampling:u40}
\end{figure}

\subparagraph*{Experimental evaluation.}
Following our considerations in \autoref{appendix:sampling-params},
\begin{enumerate}[(1)]
\item we set $b := 2^{24}$ for buckets backed by \emph{bit vectors},
\item we set $b := 2^{10}$ for buckets backed by \emph{unsorted lists} and
\item for \emph{hybrid} buckets, we set $\theta_{\min} := 2^9$ and $\theta_{\max} := 2^{10}$ and try different sizes $b$.
\end{enumerate}

Our results are shown in \autoref{fig:sampling:u40}.
We first discuss the results for 32-bit keys.
In most configurations, we achieve throughputs higher than $2^{22}$ operations per second for both updates (insertions and deletions) and predecessor queries.
Furthermore, we achieve compression in that we require less than 32 bits per key, because we only store trunacted keys within the buckets.
The compression increases with larger $n$ as for sufficiently large $n$, all buckets are active and more (truncated) keys are inserted into the same number of buckets.
We observe that the top level organization, array versus hash table, barely appears to matter.
The only difference is a slightly slower performance, but also lower memory consumption of the hash tables for smaller $n$, which was to be expected.
However, as all buckets become active for larger $n$, these differences become negligible.

We now look at the three types of bucket organization.
The clear outliers are where we implement buckets as unsorted lists of size at most $2^{10}$: here, all operations are between 2--4 times slower than the rest, and the number of inserted keys visibly affects the performance of queries and deletions negatively, which is due to linear scans facing a higher bucket fill rate.
Hybrid buckets and those backed by bit vectors appear to be on par especially for large $n$, as the hybrid representation eventually switches to bit vectors.
As expected, the bit vector representation achieves higher compression than the unsorted list representation.

Now, we discuss the results for 40-bit keys.
The memory consumption is obviously different:
while we still achieve compression below 40 bits per key for large $n$, the top level now contains up to $2^{30}$ active buckets (for buckets of size $2^{10}$), causing a big memory overhead that can only be compensated for sufficiently large $n$.
Hybrid buckets may cause an explosion of memory consumption when they switch to bit vectors, as can be seen for bucket size $2^{24}$ at $2^{30}$ keys.
For $2^{33}$ keys, we can see how it slowly starts to compensate.
Regarding performance, similar observations as for 32-bit keys can be made, except that the top level organization now does matter: the smaller the buckets,
the more linear scans weigh in, such that the hash table approach becomes faster for updates but slower for queries.
Since this data structure is not suitable for large universes, we omit experiments for 64-bit keys.

\section{Y-Fast Tries}
\label{sec:yfast}

The \emph{x-fast} trie by Willard \cite{DBLP:journals/ipl/Willard83} is conceptually a variation of the binary trie where
\begin{enumerate}[(1)]
\item the keys of $S$ are doubly-linked in ascending order and
\item if a node does not have a left (right) child, then the corresponding pointer is replaced by a \emph{descendant} pointer that directly points to the smallest (largest) leaf descending from it.
\end{enumerate}
The trie is stored in the \emph{level-search data structure} (LSS).
We say that a node of the trie is on level $\ell$ if it has depth $\ell$.
For each level $\ell$ of the trie, the LSS stores an entry for every node $v$ that exists on level $\ell$, which we identify by the bit sequence $B_v \in \{0,1\}^\ell$ that encodes the path in the trie from the root to $v$.
Specifically, the LSS associates $B_v$ to $v$'s descendant pointers.
We can find $\qpred_S(x)$ in expected time $\Oh{\lg\lg u}$ as follows:
we first binary search the $\ceil{\lg u}$ levels of the trie to find the bottom-most node $v$ on the path leading to $x$ if it were contained in $S$.
On each level $\ell$ that we inspect, we query the bit prefix $x\bitaccess{0 \twodots{} \ell-1}$ in the LSS in $\Oh{1}$ expected time to test if we are done.
From $v$, by construction, we can take a descendant pointer to the predecessor or successor of $x$.
Updates of the x-fast trie require $\Oh{\lg u}$ expected time as in the worst case, the LSS needs to be updated for every level following an insertion or deletion.
The total memory consumption of the x-fast trie is $\Oh{n \lg u}$ words.

The \emph{y-fast trie} improves this to $\Oh{n}$ words:
we partition $S$ into $\Th{n/\lg u}$ buckets of $\Th{\lg u}$ keys each and determine a \emph{representative} for each bucket, e.g., the minimum contained key.
Then, we build an x-fast trie over only the representatives, which occupies $\Oh{n}$ words of memory.
For each bucket, we construct a binary search tree for the keys contained in it, consuming $\Oh{(n/\lg u) \cdot \lg u} = \Oh{n}$ words.
When looking for the predecessor of $x$, we can locate its bucket using the x-fast trie over the representatives in expected time $\Oh{\lg\lg u}$ and within the buckets, searching and updating can be done in time $\Oh{\lg\lg u}$.
The sampling of representatives also improves the amortized expected update times to $\Oh{\lg\lg u}$.

\begin{figure}[tb]
\centering
\begin{tiny}
\begin{tikzpicture}[
    level 1/.style={sibling distance=28em},
    level 2/.style={sibling distance=14em},
    level 3/.style={sibling distance=7em},
    level 4/.style={sibling distance=3.5em},
    level 5/.style={sibling distance=1.75em},
    level distance=5ex,
    inner/.style={shape=circle,draw},
    leaf/.style={shape=rectangle,draw,minimum width=1em,minimum height=1em},
    nleaf/.style={},
    bucket/.style={draw=gray,thick,inner sep=0.175em},
]
\node [inner, ultra thick] {}
    child { node [inner, ultra thick] {}
        child { node [inner,ultra thick] {}
            child { node [inner,ultra thick] {}
                child { node [inner,ultra thick] {}
                    child { node [leaf,ultra thick] (key-0) {}}
                    child { node [nleaf] {}}
                }
                child { node [inner] {}
                    child { node [nleaf] {}}
                    child { node [leaf] (key-3) {}}
                }
            }
            child { node [inner] {}
                child { node [inner] {}
                    child { node [nleaf] {}}
                    child { node [nleaf] {}}
                }
                child { node [inner] {}
                    child { node [leaf] (key-6) {}}
                    child { node [leaf] (key-7) {}}
                }
            }
        }
        child { node [inner] {}
            child { node [inner] {}
                child { node [inner] {}
                    child { node [nleaf] {}}
                    child { node [leaf] (key-9) {}}
                }
                child { node [inner] {}
                    child { node [nleaf] {}}
                    child { node [nleaf] {}}
                }
            }
            child { node [inner] {}
                child { node [inner] {}
                    child { node [nleaf] {}}
                    child { node [nleaf] {}}
                }
                child { node [inner] {}
                    child { node [nleaf] {}}
                    child { node [nleaf] {}}
                }
            }
        }
    }
    child { node [inner, ultra thick] {}
        child { node [inner,ultra thick] {}
            child { node [inner,ultra thick] {}
                child { node [inner,ultra thick] {}
                    child { node [nleaf] {}}
                    child { node [leaf,ultra thick] (key-17) {}}
                }
                child { node [inner] {}
                    child { node [leaf] (key-18) {}}
                    child { node [leaf] (key-19) {}}
                }
            }
            child { node [inner,ultra thick] {}
                child { node [inner,ultra thick] {}
                    child { node [leaf,ultra thick] (key-20) {}}
                    child { node [leaf] (key-21) {}}
                }
                child { node [inner] {}
                    child { node [nleaf] {}}
                    child { node [leaf] (key-23) {}}
                }
            }
        }
        child { node [inner] {}
            child { node [inner] {}
                child { node [inner] {}
                    child { node [nleaf] {}}
                    child { node [nleaf] {}}
                }
                child { node [inner] {}
                    child { node [nleaf] {}}
                    child { node [nleaf] {}}
                }
            }
            child { node [inner] {}
                child { node [inner] {}
                    child { node [nleaf] {}}
                    child { node [nleaf] {}}
                }
                child { node [inner] {}
                    child { node [nleaf] {}}
                    child { node [nleaf] {}}
                }
            }
        }
    }
;

\node [bucket,fit=(key-0) (key-9)] (bucket-1) {};
\node [bucket,fit=(key-17) (key-19)] (bucket-2) {};
\node [bucket,fit=(key-20) (key-23)] (bucket-3) {};
\draw [] (key-0.north west) to (key-0.south east);
\draw [] (key-0.south west) to (key-0.north east);
\draw [] (key-20.north west) to (key-20.south east);
\draw [] (key-20.south west) to (key-20.north east);

\draw [dotted,gray] (bucket-1.east) to (bucket-2.west);
\draw [dotted,gray] (bucket-2.east) to (bucket-3.west);

\draw [dashed,gray] (-28em,-5ex) to (28em,-5ex);
\node [xshift=1em,yshift=1em] at (-28em, -5ex) {\small $\ell_\top$};

\draw [dashed,gray] (-28em,-15ex) to (28em,-15ex);
\node [xshift=1em,yshift=1em] at (-28em, -15ex) {\small $\ell_\bot$};

\node [below=0.5em of key-0.south]  {\small $B_1$};
\node [below=0.5em of key-17.south]  {\small $B_2$};
\node [below=0.5em of key-20.south]  {\small $B_3$};

\end{tikzpicture}
\end{tiny}
\caption{
    Our y-fast trie for $w=5$ and $S = \{ 3, 6, 7, 9, 17, 18, 19, 21, 23 \}$ with $t=2$, $c=2$ and $\gamma=1$.
    Edge and leaf labels of the conceptual trie are omitted for the sake of clarity:
    left edges are labeled by 0, right edges by 1 and leaves are labeled by the corresponding keys.
    Keys contained in $S$ are shown as squares, where representatives of buckets have a thick border.
    Representatives marked with an X are deleted:
    they are still representatives of buckets, but no longer contained in $S$ themselves.
    Buckets are shown as rectangles around the contained keys.
    Nodes on paths that lead to representatives are contained in the x-fast trie's LSS and are drawn with a thick border;
    other nodes are not contained in the LSS.
    Levels $\ell_\top$ and $\ell_\bot$ are highlighted by dashed lines.
}
\label{fig:yfast-trie}
\end{figure}
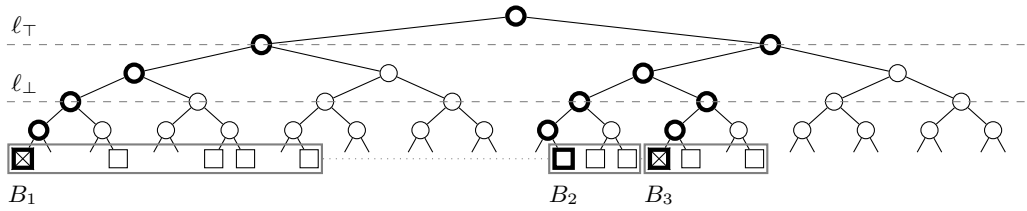

\subparagraph*{Implementation.}
Let $t = \Th{\lg u}$, $\gamma > 0$, and $c > 2\gamma$ be parameters.
We partition $S$ into buckets of size variable in $[\gamma t, ct]$.
We name the minimum key contained in a bucket its representative and only representatives are contained in the x-fast trie, which has height $\ceil{\lg u - \lg t}$ as the lowest $\lg t$ bits of the keys are maintained in the buckets.
Within a bucket, we store keys in a sorted or unsorted list rather than a binary search tree.
This increases the asymptotic time needed for updates and predecessor queries, but the additional memory costs for structures such as binary trees, albeit asymptotically constant, would be too high in practice.
\autoref{fig:yfast-trie} shows an example of our y-fast trie that we describe in the following.

We try to keep $t$ small such that buckets fit into few consecutive cache lines and can be searched quickly.
Because this directly affects the height of the x-fast trie maintaining the representatives, we speed up searches as follows:
let $\ell_\top$ be the bottommost level where \emph{all} possible nodes exist in the x-fast trie and let $\ell_\bot$ be the topmost level where \emph{no branching nodes} exist in the x-fast trie.
Consider an operation involving a key $x \in U$:
we locate its bucket by a vertical binary search in the x-fast trie's LSS.
Because all levels above $\ell_\top$ contain all possible nodes and because all nodes on levels below $\ell_\bot$ point to the same buckets as their respective ancestors on level $\ell_\bot$, we limit the binary search to the levels between $\ell_\top$ and $\ell_\bot$ and maintain $\ell_\bot$ and $\ell_\top$ under updates with no asymptotic extra cost.
With this strategy, we can also save space by avoiding storage of any nodes on levels below $\ell_\bot$;
the corresponding hash tables in the LSS simply remain empty.
Intuitively, this cuts off trailing unary paths in the x-fast trie.
Note that due to the sampling mechanism, we always have $\ell_\bot \le \lg(\gamma t)$, so the $\Th{\lg \lg u}$ bottommost levels are never stored.

To speed up deletions, we allow the representative of a bucket to be no longer contained in the bucket by itself.
When it is deleted, we mark it as such, but it remains the bucket's representative and also remains in the x-fast trie.
This strategy avoids the need of finding a new representative and updating the x-fast trie every time a representative is deleted.
However, we must consider a special case when answering predecessor queries.
Let $y_\text{rep}$ be the deleted representative of a bucket and $y_\text{min} > y_\text{rep}$ the smallest key currently contained in the bucket,
and consider the query $\qpred_S(x)$ with $y_\text{rep} \leq x < y_\text{min}$.
The x-fast trie will lead us to said bucket and $y_\text{rep}$ would be the predecessor of $x$.
When we detect $y_\text{rep}$ as deleted, we follow a pointer to the preceding bucket, which must contain the predecessor of $x$.

Buckets are merged and split as in B-trees \cite[chapter~18]{DBLP:books/daglib/0023376} to ensure their size stays within $[\gamma t, ct]$.
To avoid the creation of a new bucket each time a new minimum is inserted into the data structure, we maintain a special bucket with representative $-\infty$ that we allow to become empty and will never be removed by a merge.

When using unsorted lists to maintain keys within a bucket, we can amortize insertion costs.
Unless a split is required, inserting a key into a bucket simply means appending it in constant time.
Thus, if we choose $c := \gamma+\Th{t} > 2\gamma$, we can amortize the time needed for a split over $\Th{t}$ constant-time insertions.
This amortization leads to $\Oh{(\lg u)/t} = \Oh{1}$ time needed for inserting a key into a bucket followed by a potential split, such that the amortized expected insertion time of the y-fast trie is $\Oh{\lg\lg u}$ as in the original.
This cannot be achieved for deletions, as the key to be deleted needs to be located in time $\Oh{\lg u}$ first.

\begin{example}[insertion]
\label{example:yfast-insert}
Consider the y-fast trie in \autoref{fig:yfast-trie} with $t=2$ and $c=2$.
We insert the new key 8.
The binary search in the x-fast trie's LSS is constrained only to the three levels between $\ell_\top$ and $\ell_\bot$ and leads to bucket $B_1$ with (deleted) representative 0.
We insert $x$ by appending it to the unsorted list of keys.
However, we then have $|B_1| = 5 > 4 = ct$, thus we have to split $B_1$.
We create a new bucket $B_1'$ with representative 7 (the median) and move keys such that $B_1 := \{ 3, 6 \}$ and $B_1' := \{ 7, 8, 9 \}$.
Even though 0 is no longer contained in $B_1$, it remains its representative.
Finally, we enter key 7 into the x-fast trie, causing two new nodes to be added to the LSS.
However, $\ell_\bot$ remains unchanged, as the newly added nodes form a unary path beginning at level $\ell_\bot$.
\end{example}

\begin{example}[deletion]
Consider the y-fast trie in \autoref{fig:yfast-trie} with $t=2$ and $\gamma=1$.
We delete key 21, which we find in bucket $B_3$ as in \autoref{example:yfast-insert}.
After deletion, we have $|B_3| = 1 < 2 = \gamma t$ (note how 20 is the representative, but is marked as deleted), thus we have to merge.
As the only neighbour, we merge with bucket $B_2$ by moving key 23 such that $B_2 := \{ 17, 18, 19, 23 \}$.
The former representative of $B_3$, key 20, is now removed from the x-fast trie.
Observe how the path of nodes leading to $B_2$ now becomes a unary path starting at level $\ell_\top$.
Because all unary paths then start at level $\ell_\top$, we set $\ell_\bot := \ell_\top$.
\end{example}

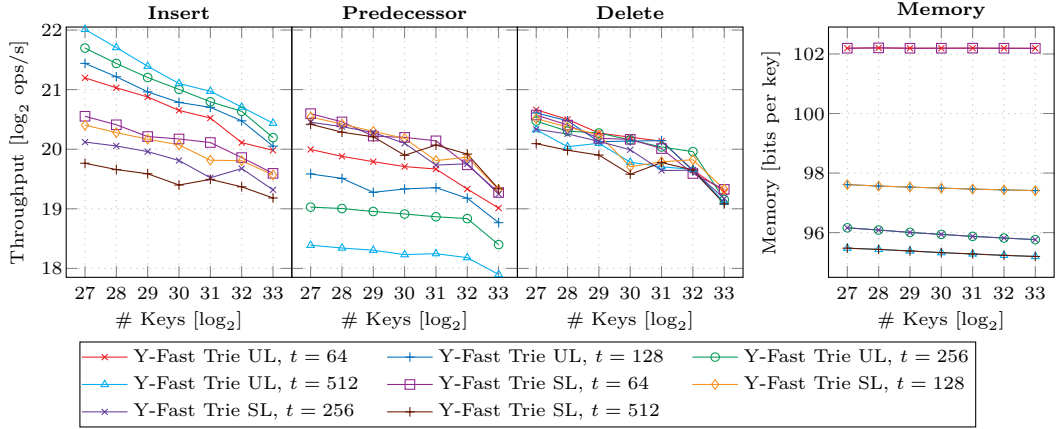
\begin{figure}[tb]
\begin{tikzpicture}
\begin{groupplot}[group3, ylogtput, ymin=17.85, ymax=22.05]
\nextgroupplot[xlognum, title={Insert}, legend to name=leg:yfast:u64]
\addplot coordinates { (27.0,21.197) (28.0,21.0305) (29.0,20.8768) (30.0,20.6511) (31.0,20.5234) (32.0,20.1107) (33.0,19.9807) };
\addlegendentry{Y-Fast Trie UL, $t=64$};
\addplot coordinates { (27.0,21.4405) (28.0,21.2157) (29.0,20.9622) (30.0,20.7855) (31.0,20.7031) (32.0,20.4764) (33.0,20.0509) };
\addlegendentry{Y-Fast Trie UL, $t=128$};
\addplot coordinates { (27.0,21.6974) (28.0,21.4393) (29.0,21.203) (30.0,21.0038) (31.0,20.7956) (32.0,20.6357) (33.0,20.1941) };
\addlegendentry{Y-Fast Trie UL, $t=256$};
\addplot coordinates { (27.0,22.0104) (28.0,21.7052) (29.0,21.3922) (30.0,21.1016) (31.0,20.9727) (32.0,20.7067) (33.0,20.4358) };
\addlegendentry{Y-Fast Trie UL, $t=512$};
\addplot coordinates { (27.0,20.5526) (28.0,20.4101) (29.0,20.2124) (30.0,20.1722) (31.0,20.1122) (32.0,19.8597) (33.0,19.5953) };
\addlegendentry{Y-Fast Trie SL, $t=64$};
\addplot coordinates { (27.0,20.4038) (28.0,20.2787) (29.0,20.168) (30.0,20.0729) (31.0,19.8155) (32.0,19.8066) (33.0,19.5686) };
\addlegendentry{Y-Fast Trie SL, $t=128$};
\addplot coordinates { (27.0,20.12) (28.0,20.0544) (29.0,19.96) (30.0,19.8103) (31.0,19.5206) (32.0,19.6756) (33.0,19.3183) };
\addlegendentry{Y-Fast Trie SL, $t=256$};
\addplot coordinates { (27.0,19.7642) (28.0,19.6586) (29.0,19.5879) (30.0,19.3989) (31.0,19.4937) (32.0,19.3687) (33.0,19.1816) };
\addlegendentry{Y-Fast Trie SL, $t=512$};
\nextgroupplot[xlognum, title={Predecessor}]
\addplot coordinates { (27.0,19.9959) (28.0,19.88) (29.0,19.7911) (30.0,19.7086) (31.0,19.6672) (32.0,19.3315) (33.0,19.0126) };
\addlegendentry{Y-Fast Trie UL, $t=64$};
\addplot coordinates { (27.0,19.5845) (28.0,19.5108) (29.0,19.2758) (30.0,19.3336) (31.0,19.3546) (32.0,19.1778) (33.0,18.7671) };
\addlegendentry{Y-Fast Trie UL, $t=128$};
\addplot coordinates { (27.0,19.0283) (28.0,19.0045) (29.0,18.9547) (30.0,18.913) (31.0,18.8665) (32.0,18.8353) (33.0,18.399) };
\addlegendentry{Y-Fast Trie UL, $t=256$};
\addplot coordinates { (27.0,18.3885) (28.0,18.3423) (29.0,18.3045) (30.0,18.2306) (31.0,18.2471) (32.0,18.1801) (33.0,17.8983) };
\addlegendentry{Y-Fast Trie UL, $t=512$};
\addplot coordinates { (27.0,20.5971) (28.0,20.4586) (29.0,20.2201) (30.0,20.1988) (31.0,20.1411) (32.0,19.7377) (33.0,19.2734) };
\addlegendentry{Y-Fast Trie SL, $t=64$};
\addplot coordinates { (27.0,20.5411) (28.0,20.4216) (29.0,20.2989) (30.0,20.1817) (31.0,19.8082) (32.0,19.8666) (33.0,19.3234) };
\addlegendentry{Y-Fast Trie SL, $t=128$};
\addplot coordinates { (27.0,20.4521) (28.0,20.3795) (29.0,20.2792) (30.0,20.0996) (31.0,19.7338) (32.0,19.7553) (33.0,19.2471) };
\addlegendentry{Y-Fast Trie SL, $t=256$};
\addplot coordinates { (27.0,20.4192) (28.0,20.2852) (29.0,20.2079) (30.0,19.8981) (31.0,20.0696) (32.0,19.918) (33.0,19.3337) };
\addlegendentry{Y-Fast Trie SL, $t=512$};
\legend{}
\nextgroupplot[xlognum, title={Delete}]
\addplot coordinates { (27.0,20.6625) (28.0,20.5005) (29.0,20.2716) (30.0,20.2014) (31.0,20.1345) (32.0,19.6502) (33.0,19.2865) };
\addlegendentry{Y-Fast Trie UL, $t=64$};
\addplot coordinates { (27.0,20.6077) (28.0,20.4776) (29.0,20.1251) (30.0,20.1373) (31.0,20.144) (32.0,19.6356) (33.0,19.1011) };
\addlegendentry{Y-Fast Trie UL, $t=128$};
\addplot coordinates { (27.0,20.4717) (28.0,20.3033) (29.0,20.272) (30.0,20.1557) (31.0,20.0368) (32.0,19.9596) (33.0,19.1557) };
\addlegendentry{Y-Fast Trie UL, $t=256$};
\addplot coordinates { (27.0,20.3272) (28.0,20.0423) (29.0,20.0991) (30.0,19.7833) (31.0,19.7046) (32.0,19.6611) (33.0,19.1084) };
\addlegendentry{Y-Fast Trie UL, $t=512$};
\addplot coordinates { (27.0,20.5733) (28.0,20.4057) (29.0,20.1833) (30.0,20.1651) (31.0,20.0109) (32.0,19.5919) (33.0,19.3259) };
\addlegendentry{Y-Fast Trie SL, $t=64$};
\addplot coordinates { (27.0,20.5132) (28.0,20.3728) (29.0,20.2407) (30.0,19.7065) (31.0,19.7888) (32.0,19.8227) (33.0,19.3281) };
\addlegendentry{Y-Fast Trie SL, $t=128$};
\addplot coordinates { (27.0,20.333) (28.0,20.2566) (29.0,20.1318) (30.0,19.9902) (31.0,19.6435) (32.0,19.6469) (33.0,19.1705) };
\addlegendentry{Y-Fast Trie SL, $t=256$};
\addplot coordinates { (27.0,20.0935) (28.0,19.9827) (29.0,19.9014) (30.0,19.5794) (31.0,19.7758) (32.0,19.6512) (33.0,19.0819) };
\addlegendentry{Y-Fast Trie SL, $t=512$};
\legend{}
\end{groupplot}
\end{tikzpicture}
\begin{tikzpicture}
\begin{axis}[plot4col, xlognum, ybitsperkey, title={Memory}]
\addplot coordinates { (27.0,102.193) (28.0,102.204) (29.0,102.192) (30.0,102.193) (31.0,102.195) (32.0,102.192) (33.0,102.189) };
\addlegendentry{Y-Fast Trie UL, $t=64$};
\addplot coordinates { (27.0,97.6131) (28.0,97.5644) (29.0,97.5294) (30.0,97.4956) (31.0,97.463) (32.0,97.4368) (33.0,97.4121) };
\addlegendentry{Y-Fast Trie UL, $t=128$};
\addplot coordinates { (27.0,96.1625) (28.0,96.0882) (29.0,96.0077) (30.0,95.9415) (31.0,95.8748) (32.0,95.8196) (33.0,95.766) };
\addlegendentry{Y-Fast Trie UL, $t=256$};
\addplot coordinates { (27.0,95.4788) (28.0,95.437) (29.0,95.3846) (30.0,95.3288) (31.0,95.2842) (32.0,95.2382) (33.0,95.2001) };
\addlegendentry{Y-Fast Trie UL, $t=512$};
\addplot coordinates { (27.0,102.193) (28.0,102.204) (29.0,102.192) (30.0,102.193) (31.0,102.195) (32.0,102.192) (33.0,102.189) };
\addlegendentry{Y-Fast Trie SL, $t=64$};
\addplot coordinates { (27.0,97.6131) (28.0,97.5644) (29.0,97.5294) (30.0,97.4956) (31.0,97.463) (32.0,97.4368) (33.0,97.4121) };
\addlegendentry{Y-Fast Trie SL, $t=128$};
\addplot coordinates { (27.0,96.1625) (28.0,96.0882) (29.0,96.0077) (30.0,95.9415) (31.0,95.8748) (32.0,95.8196) (33.0,95.766) };
\addlegendentry{Y-Fast Trie SL, $t=256$};
\addplot coordinates { (27.0,95.4788) (28.0,95.437) (29.0,95.3846) (30.0,95.3288) (31.0,95.2842) (32.0,95.2382) (33.0,95.2001) };
\addlegendentry{Y-Fast Trie SL, $t=512$};
\legend{}
\end{axis}
\end{tikzpicture}
\centering\ref{leg:yfast:u64}
\caption{
Throughputs for the insert, predecessor and delete operations, as well as memory usage of the y-fast trie for $U=[2^{64}]$.
Best viewed in colour.
\emph{UL} stands for unsorted, \emph{SL} for sorted lists.
}
\label{fig:yfast:u64}
\end{figure}

\subparagraph*{Experimental evaluation.}
In our experiments, we set $c:=2$ and $\gamma:=1/4$ and choose $t$ as powers of two to optimize memory alignments.
Our results for $t := 64$ to $512$ and 64-bit keys are shown in \autoref{fig:yfast:u64}.
(Additional results are given in \autoref{fig:yfast:u40} in \autoref{appendix:results}.)

The fastest predecessor queries are achieved when buckets are organized as sorted lists and binary search is used to answer bucket-level queries.
Conversely, insertions are are up to twice as fast in the unsorted list case, where new keys simply have to be appended without preserving any order.
Regarding deletions, there is no substantial difference between using a sorted or unsorted list to organize the buckets: while we can find the item to be deleted using binary search when using a sorted list, we have to shift up to $t$ keys afterwards.
As we use simple arrays for storage in either case, there is also no difference in memory consumption.

As expected, the bucket size of $t$ is a direct trade-off parameter for update versus query performance and memory usage, which is very visible when buckets are organized as unsorted lists.
Here, insertions become faster as the bucket size grows since they are trivial on the bucket level and the LSS needs to be updated less often.
However, larger buckets mean longer scans when answering predecessor queries.
The bucket size is much less impactful on query performance when sorted lists are used, as the bucket-level query time is then only logarithmic in the bucket size.
The memory consumption is also affected by the bucket size: larger buckets imply less levels in the LSS and thus less memory needed.


As a conclusion, unsorted lists may be preferable when fast insertions are required and the performance of predecessor queries is less important.
For the general case, however, using sorted lists appears to be preferable, as all operations then have similar throughputs.

\begin{figure}[t]
\centering
\begin{subfigure}[b]{0.47\textwidth}
\centering
\begin{small}
\begin{tikzpicture}[
    sibling distance=5em,
    level distance=5.5ex,
    inner/.style = {shape=circle,draw},
    leaf/.style = {shape=rectangle,draw},
]
\node [inner,ultra thick] {}
    child { node [inner,ultra thick] {}
        child { node [inner] {}
            child { node [inner] {}
                child { node [inner, ultra thick] {}
                    child { node [leaf] {2} edge from parent node [left] {\textbf{0}} }
                    child { node [leaf] {3} edge from parent node [right] {\textbf{1}} }
                    edge from parent node [left] {1}
                }
                edge from parent node [left] {0}
            }
            edge from parent node [left] {\textbf{0}}
        }
        child { node [inner] {}
            child { node [inner] {}
                child { node [inner] {}
                    child { node [leaf] {12} edge from parent node [right] {\textbf{0}} }
                    edge from parent node [right] {0}
                }
                edge from parent node [right] {1}
            }
            edge from parent node [right] {\textbf{1}}
        }
        edge from parent node [left] {\textbf{0}}
    }
    child { node [inner] {}
        child { node [inner] {}
            child { node [inner] {}
                child { node [inner] {}
                    child { node [leaf] {27} edge from parent node [right] {\textbf{1}} }
                    edge from parent node [right] {1}
                }
                edge from parent node [right] {0}
            }
            edge from parent node [right] {\textbf{1}}
        }
        edge from parent node [right] {\textbf{1}}
    }
;
\end{tikzpicture}
\end{small}
\caption{
    The binary trie for $S$.
    Branching nodes have thicker outlines and bits at distinguishing positions are written in bold.
}
\label{fig:trie}
\end{subfigure}
\hspace{0.04\textwidth}
\begin{subfigure}[b]{0.47\textwidth}
\centering
\begin{small}
$M=\texttt{11001}$\\
\vspace{0.5ex}
\begin{tabular}{|r|c|c|c|c|c|}
\toprule
$x$ & \scriptsize{binary} & $\hat{x}$ & $\hat{x}^{\bitdontcare}$ & \scriptsize{\wbranch{}} & \scriptsize{\wfree{}} \\
\midrule
2 & \texttt{00010} & \texttt{000} & \bitzero\bitzero\bitzero & \texttt{000} & \texttt{000} \\
3 & \texttt{00011} & \texttt{001} & \bitzero\bitzero\bitone & \texttt{001} & \texttt{000} \\
12 & \texttt{01100} & \texttt{010} & \bitzero\bitone\bitdontcare & \texttt{010} & \texttt{001} \\
27 & \texttt{11011} & \texttt{111} & \bitone\bitdontcare\bitdontcare & \texttt{100} & \texttt{011} \\
\bottomrule
\end{tabular}
\end{small}
\caption{
    The keys $x \in S$, with binary representation and compressed versions $\hat{x}$ and $\hat{x}^{\bitdontcare}$ without and with don't cares according to \cite{DBLP:journals/jcss/FredmanW93} and \cite{DBLP:conf/focs/PatrascuT14}, respectively.
    \wbranch{} and \wfree{} encode the matrix given by column $\hat{x}^{\bitdontcare}$ as described in \cite{DBLP:conf/focs/PatrascuT14}.
    The mask $M$ marks the distinguishing positions of $S$.
}
\label{fig:compressed-keys}
\end{subfigure}
\caption{
    Binary trie and compressed keys for $S=\{2,3,12,27\}$ and $w = 5$.
}
\label{fig:dynamic-fusion-node}
\end{figure}

\section{Fusion Trees}
\label{sec:fusion}

\Patrascu{} and Thorup \cite{DBLP:conf/focs/PatrascuT14} introduce \emph{dynamic fusion nodes} as a sorted list data structure for $|S| \leq k \leq \sqrt{w}$ keys that supports predecessor queries and updates in time $\Oh{1}$.
It is based on the fusion node, originally described by Fredman and Willard \cite{DBLP:journals/jcss/FredmanW93}, that simulates navigation in a compact binary trie of $S$ represented by compressed keys.
Given a key $x \in S$, we only consider those bits at \emph{distinguishing} positions.
A position $\ell < \lceil\lg u\rceil$ is a distinguishing position if there is at least one branch on level $\ell$ in the binary trie.
For $k$ keys, there can be at most $k$ branches in the binary trie, and thus there can be at most $k$ distinguishing positions.
A \emph{compressed key} $\hat{x}$ consists of only the bits of $x$ at distinguishing positions moved to the $k$ least significant positions.
We maintain the set of distinguishing positions in a \emph{mask} $M$ of $w$ bits where the $k$ distinguishing bits are set and all other bits are clear.
\autoref{fig:compressed-keys} shows an example.
From $x$, we can compute $\hat{x}$ in constant time by masking out unwanted bits using $M$, followed by multiplications to relocate the distinguishing bits.
The $k$ compressed keys of $S$ can be stored in a $k \times k$ bit matrix $\hat{S}$ that fits into a single word.
With this, we can compute $\qpred_S(x)$ in time $\Oh{1}$ as Fredman and Willard describe in \cite{DBLP:journals/jcss/FredmanW93}.
However, updates may cause a new position to become distinguishing after an insertion, or a position to be no longer distinguishing after a deletion.
In these cases, their data structure needs to be rebuilt from scratch.
To resolve this, \Patrascu{} and Thorup introduce \emph{don't care} bits (written \bitdontcare) that indicate bits at distinguishing position that are, however, not used for branching in a specific compressed key.
The data structure now contains a $k \times k$ matrix over the new alphabet $\{\bitzero, \bitone, \bitdontcare\}$, which we encode using two $k \times k$ bit matrices that fit into one word each.
Examples for this can be seen in \autoref{fig:compressed-keys}.
The notion of wildcards allows for updating the data structure in time $\Oh{1}$.
We refer to \autoref{appendix:fusion} for a more detailed description and examples, including an elaboration of the deletion of keys not given in \cite{DBLP:conf/focs/PatrascuT14}.

A \emph{B-tree} is a self-balancing multiary tree data structure for representing a dynamic ordered set of items.
With $B$ the maximum degree of a node, it is guaranteed to maintain height $\log_B n$, such that lookup---including predecessor---queries and updates can be done in time $\Oh{\log_B n}$.
We consider B-trees a well-known folklore data structure and refer to \cite[chapter~18]{DBLP:books/daglib/0023376} for a comprehensive introduction.
Embedding fusion nodes into a B-tree, using the keys contained in the nodes as splitters, are the typical ingredients of a \emph{fusion tree}.

\subparagraph*{Implementation.}
As we deal with 64-bit architectures ($w=64$), we choose $k:=8$, such that a $k \times k$ bit matrix can be stored in a single word represented row-wise by an array $\hat{X} = [\hat{x}_0, \dots, \hat{x}_7]$ of compressed keys.
We keep $\hat{X}$ in ascending order, i.e., $\hat{x}_0 < \hat{x}_1 < \dots < \hat{x}_7$.

The most important operation is key compression, writing only the bits of $x$ at distinguishing positions into a word $\hat{x}$.
Instead of an approach based on sparse tables \cite{DBLP:journals/cacm/TarjanY79}, we make use of the \emph{parallel bits extract} (\pext) instruction \cite{intel-instruction-set-reference}.
Let $M$ be the $w$-bit mask identifying distinguishing positions.
Then, conveniently, $\hat{x} = \pext(x, M)$.
Another core operation is finding the rank $i$ of a compressed key $\hat{y}$ in $\hat{X}$.
For this, we use the MMX SIMD instruction \pcmpgtb\ \cite{intel-instruction-set-reference}, which performs a byte-wise greater-than comparison of two 64-bit words.
First, we multiply $\hat{y}$ by the constant $(0^{k-1}1)^k$ to retrieve the word $\hat{y}^k$ containing $k$ copies of $\hat{y}$.
Let $j$ be the smallest rank such that $\hat{x}_j > \hat{y}$.
The instruction $\pcmpgtb(\hat{X}, \hat{y}^k)$ returns the word $B_{\hat{X}>\hat{y}}$ where the $kj$ lowest bits are zero and the remaining bits are set (because $\hat{X}$ is ordered).
Therefore, $j = \lfloor\tzcnt(B_{\hat{X}>\hat{y}}) / k\rfloor$ and finally $i = j-1$.
Alternatively, because $\hat{X}$ easily fits into a cache line, we consider a na\"ive linear search.

We extend our implementation to support also $k=16$ by simulating a 256-bit word using four 64-bit words.
The special CPU instructions can be extended by executing them on each of the four 64-bit words and then combining the results.
The processors to our disposal actually support a variant of \pcmpgtb\ for the parallel comparison of sixteen 16-bit words contained in a 256-bit word (namely the Intel intrisic \texttt{\_mm256\_cmpgt\_epi16}).

We implement B-trees largely following the description in \cite[chapter~18]{DBLP:books/daglib/0023376}.
Nodes have at most $B$ children and thus contain up to $B-1$ keys used as splitters.
When plugging fusion nodes into B-trees of degree $k$, we have \emph{fusion trees}.

\begin{figure}[tb]
\begin{tikzpicture}
\begin{groupplot}[group3, ylogtput, ymin=17.5, ymax=20.675]
\nextgroupplot[xlognum, title={Insert}, legend to name=leg:fusion:u64]
\addplot coordinates { (27.0,20.2047) (28.0,20.0683) (29.0,19.9918) (30.0,19.8965) (31.0,19.7041) (32.0,19.5991) (33.0,19.4226) };
\addlegendentry{B-tree LS, $B=128$};
\addplot coordinates { (27.0,20.4053) (28.0,20.2324) (29.0,20.0297) (30.0,19.7638) (31.0,19.6176) (32.0,19.4563) (33.0,18.9688) };
\addlegendentry{B-tree LS, $B=16$};
\addplot coordinates { (27.0,20.0097) (28.0,19.8413) (29.0,19.6169) (30.0,19.5373) (31.0,19.4032) (32.0,19.3234) (33.0,19.0801) };
\addlegendentry{B-tree LS, $B=256$};
\addplot coordinates { (27.0,20.5038) (28.0,20.3444) (29.0,20.2123) (30.0,19.8634) (31.0,19.9013) (32.0,19.7776) (33.0,19.42) };
\addlegendentry{B-tree LS, $B=64$};
\addplot coordinates { (27.0,20.0624) (28.0,19.8891) (29.0,19.677) (30.0,19.4967) (31.0,19.2329) (32.0,18.9405) (33.0,18.5039) };
\addlegendentry{B-tree LS, $B=8$};
\addplot coordinates { (27.0,20.0484) (28.0,19.8147) (29.0,19.7047) (30.0,19.6501) (31.0,19.4663) (32.0,19.2944) (33.0,19.1115) };
\addlegendentry{B-tree BS, $B=128$};
\addplot coordinates { (27.0,20.1105) (28.0,19.903) (29.0,19.7783) (30.0,19.6694) (31.0,19.4256) (32.0,19.2126) (33.0,18.872) };
\addlegendentry{B-tree BS, $B=16$};
\addplot coordinates { (27.0,19.854) (28.0,19.7012) (29.0,19.4542) (30.0,19.3037) (31.0,19.2313) (32.0,19.1503) (33.0,18.8966) };
\addlegendentry{B-tree BS, $B=256$};
\addplot coordinates { (27.0,20.1002) (28.0,19.9829) (29.0,19.7559) (30.0,19.8265) (31.0,19.4875) (32.0,19.4553) (33.0,19.2941) };
\addlegendentry{B-tree BS, $B=64$};
\addplot coordinates { (27.0,20.0147) (28.0,19.8284) (29.0,19.5291) (30.0,19.4496) (31.0,19.1988) (32.0,18.926) (33.0,18.5431) };
\addlegendentry{B-tree BS, $B=8$};
\addplot coordinates { (27.0,19.0079) (28.0,18.9125) (29.0,18.7732) (30.0,18.6545) (31.0,18.5859) (32.0,18.3837) (33.0,18.0566) };
\addlegendentry{Fusion Tree SIMD, $k=16$};
\addplot coordinates { (27.0,19.345) (28.0,19.175) (29.0,19.0194) (30.0,18.7821) (31.0,18.6291) (32.0,18.4518) (33.0,18.0423) };
\addlegendentry{Fusion Tree SIMD, $k=8$};
\addplot coordinates { (27.0,19.1415) (28.0,19.0253) (29.0,18.8407) (30.0,18.7979) (31.0,18.6585) (32.0,18.5546) (33.0,18.1304) };
\addlegendentry{Fusion Tree LS, $k=16$};
\addplot coordinates { (27.0,19.6335) (28.0,19.4594) (29.0,19.274) (30.0,19.0757) (31.0,18.9225) (32.0,18.5564) (33.0,18.2886) };
\addlegendentry{Fusion Tree LS, $k=8$};
\nextgroupplot[xlognum, title={Predecessor}]
\addplot coordinates { (27.0,20.1001) (28.0,20.1945) (29.0,20.0995) (30.0,20.0142) (31.0,19.8055) (32.0,19.7235) (33.0,19.2309) };
\addlegendentry{B-tree LS, $B=128$};
\addplot coordinates { (27.0,20.1647) (28.0,19.9606) (29.0,19.7151) (30.0,19.5013) (31.0,19.1773) (32.0,18.9013) (33.0,18.435) };
\addlegendentry{B-tree LS, $B=16$};
\addplot coordinates { (27.0,20.2455) (28.0,20.0478) (29.0,19.8496) (30.0,19.718) (31.0,19.1959) (32.0,19.582) (33.0,19.0151) };
\addlegendentry{B-tree LS, $B=256$};
\addplot coordinates { (27.0,20.5543) (28.0,20.3666) (29.0,20.2073) (30.0,19.7425) (31.0,19.9058) (32.0,19.7869) (33.0,19.1477) };
\addlegendentry{B-tree LS, $B=64$};
\addplot coordinates { (27.0,19.8158) (28.0,19.5939) (29.0,19.3719) (30.0,19.206) (31.0,19.0186) (32.0,18.3179) (33.0,17.9476) };
\addlegendentry{B-tree LS, $B=8$};
\addplot coordinates { (27.0,19.9706) (28.0,19.7329) (29.0,19.5821) (30.0,19.7234) (31.0,19.5668) (32.0,19.3416) (33.0,18.9741) };
\addlegendentry{B-tree BS, $B=128$};
\addplot coordinates { (27.0,19.7894) (28.0,19.4347) (29.0,19.4432) (30.0,19.5026) (31.0,19.2147) (32.0,18.9729) (33.0,18.4571) };
\addlegendentry{B-tree BS, $B=16$};
\addplot coordinates { (27.0,19.9776) (28.0,19.7961) (29.0,19.4333) (30.0,19.3411) (31.0,19.2481) (32.0,19.2655) (33.0,18.8294) };
\addlegendentry{B-tree BS, $B=256$};
\addplot coordinates { (27.0,19.7552) (28.0,19.7717) (29.0,19.5683) (30.0,19.6881) (31.0,19.2968) (32.0,19.42) (33.0,18.8948) };
\addlegendentry{B-tree BS, $B=64$};
\addplot coordinates { (27.0,19.6718) (28.0,19.4422) (29.0,19.0761) (30.0,19.2377) (31.0,19.0645) (32.0,18.3242) (33.0,18.0614) };
\addlegendentry{B-tree BS, $B=8$};
\addplot coordinates { (27.0,19.4016) (28.0,19.2581) (29.0,19.0395) (30.0,18.9312) (31.0,18.8554) (32.0,18.3181) (33.0,17.9143) };
\addlegendentry{Fusion Tree SIMD, $k=16$};
\addplot coordinates { (27.0,19.3166) (28.0,19.1124) (29.0,18.9265) (30.0,18.74) (31.0,18.6007) (32.0,17.9737) (33.0,17.6378) };
\addlegendentry{Fusion Tree SIMD, $k=8$};
\addplot coordinates { (27.0,19.534) (28.0,19.357) (29.0,19.0836) (30.0,19.0439) (31.0,18.9244) (32.0,18.427) (33.0,17.9772) };
\addlegendentry{Fusion Tree LS, $k=16$};
\addplot coordinates { (27.0,19.686) (28.0,19.2758) (29.0,19.2515) (30.0,19.0622) (31.0,18.8409) (32.0,18.2902) (33.0,17.9662) };
\addlegendentry{Fusion Tree LS, $k=8$};
\legend{}
\nextgroupplot[xlognum, title={Delete}]
\addplot coordinates { (27.0,19.6816) (28.0,19.6885) (29.0,19.4324) (30.0,19.6514) (31.0,19.4734) (32.0,19.0846) (33.0,18.9778) };
\addlegendentry{B-tree LS, $B=128$};
\addplot coordinates { (27.0,20.0721) (28.0,19.8612) (29.0,19.6426) (30.0,19.131) (31.0,19.1585) (32.0,18.8891) (33.0,18.4665) };
\addlegendentry{B-tree LS, $B=16$};
\addplot coordinates { (27.0,19.6237) (28.0,19.4215) (29.0,19.3882) (30.0,19.1432) (31.0,19.0391) (32.0,19.1984) (33.0,18.6316) };
\addlegendentry{B-tree LS, $B=256$};
\addplot coordinates { (27.0,20.2638) (28.0,19.949) (29.0,19.9481) (30.0,19.549) (31.0,19.6834) (32.0,19.4257) (33.0,18.8265) };
\addlegendentry{B-tree LS, $B=64$};
\addplot coordinates { (27.0,19.7618) (28.0,19.5386) (29.0,19.2912) (30.0,18.9443) (31.0,18.8376) (32.0,18.3248) (33.0,18.0302) };
\addlegendentry{B-tree LS, $B=8$};
\addplot coordinates { (27.0,19.8065) (28.0,19.6003) (29.0,19.4667) (30.0,19.4481) (31.0,19.2694) (32.0,19.12) (33.0,18.751) };
\addlegendentry{B-tree BS, $B=128$};
\addplot coordinates { (27.0,19.7454) (28.0,19.4601) (29.0,19.3306) (30.0,19.2134) (31.0,19.0903) (32.0,18.8543) (33.0,18.3751) };
\addlegendentry{B-tree BS, $B=16$};
\addplot coordinates { (27.0,19.6334) (28.0,19.4764) (29.0,19.1647) (30.0,19.0466) (31.0,18.9749) (32.0,18.9962) (33.0,18.5785) };
\addlegendentry{B-tree BS, $B=256$};
\addplot coordinates { (27.0,19.7932) (28.0,19.6969) (29.0,19.3448) (30.0,19.4375) (31.0,19.0873) (32.0,19.2371) (33.0,18.6845) };
\addlegendentry{B-tree BS, $B=64$};
\addplot coordinates { (27.0,19.703) (28.0,19.4152) (29.0,19.1521) (30.0,19.0887) (31.0,18.8537) (32.0,18.2198) (33.0,17.9781) };
\addlegendentry{B-tree BS, $B=8$};
\addplot coordinates { (27.0,18.6327) (28.0,18.5184) (29.0,18.4059) (30.0,18.3088) (31.0,18.0785) (32.0,17.774) (33.0,17.5537) };
\addlegendentry{Fusion Tree SIMD, $k=16$};
\addplot coordinates { (27.0,18.9912) (28.0,18.8079) (29.0,18.6466) (30.0,18.4216) (31.0,18.2526) (32.0,17.844) (33.0,17.5296) };
\addlegendentry{Fusion Tree SIMD, $k=8$};
\addplot coordinates { (27.0,18.7434) (28.0,18.6271) (29.0,18.4531) (30.0,18.3638) (31.0,18.1394) (32.0,17.9815) (33.0,17.615) };
\addlegendentry{Fusion Tree LS, $k=16$};
\addplot coordinates { (27.0,19.1997) (28.0,18.9063) (29.0,18.5786) (30.0,18.6071) (31.0,18.5537) (32.0,18.0467) (33.0,17.766) };
\addlegendentry{Fusion Tree LS, $k=8$};
\legend{}
\end{groupplot}
\end{tikzpicture}
\begin{tikzpicture}
\begin{axis}[plot4col, xlognum, ybitsperkey, title={Memory}]
\addplot coordinates { (27.0,94.3889) (28.0,94.3828) (29.0,94.374) (30.0,94.3742) (31.0,94.3738) (32.0,94.3738) (33.0,94.3749) };
\addlegendentry{B-tree LS, $B=128$};
\addplot coordinates { (27.0,116.081) (28.0,116.081) (29.0,116.081) (30.0,116.085) (31.0,116.083) (32.0,116.083) (33.0,116.083) };
\addlegendentry{B-tree LS, $B=16$};
\addplot coordinates { (27.0,93.8606) (28.0,93.8069) (29.0,93.7572) (30.0,93.7213) (31.0,93.6901) (32.0,93.657) (33.0,93.6302) };
\addlegendentry{B-tree LS, $B=256$};
\addplot coordinates { (27.0,96.4883) (28.0,96.487) (29.0,96.4839) (30.0,96.4889) (31.0,96.4865) (32.0,96.4905) (33.0,96.4922) };
\addlegendentry{B-tree LS, $B=64$};
\addplot coordinates { (27.0,131.216) (28.0,131.209) (29.0,131.216) (30.0,131.214) (31.0,131.215) (32.0,131.214) (33.0,131.215) };
\addlegendentry{B-tree LS, $B=8$};
\addplot coordinates { (27.0,94.3889) (28.0,94.3828) (29.0,94.374) (30.0,94.3742) (31.0,94.3738) (32.0,94.3738) (33.0,94.3749) };
\addlegendentry{B-tree BS, $B=128$};
\addplot coordinates { (27.0,116.081) (28.0,116.081) (29.0,116.081) (30.0,116.085) (31.0,116.083) (32.0,116.083) (33.0,116.083) };
\addlegendentry{B-tree BS, $B=16$};
\addplot coordinates { (27.0,93.8606) (28.0,93.8069) (29.0,93.7572) (30.0,93.7213) (31.0,93.6901) (32.0,93.657) (33.0,93.6302) };
\addlegendentry{B-tree BS, $B=256$};
\addplot coordinates { (27.0,96.4883) (28.0,96.487) (29.0,96.4839) (30.0,96.4889) (31.0,96.4865) (32.0,96.4905) (33.0,96.4922) };
\addlegendentry{B-tree BS, $B=64$};
\addplot coordinates { (27.0,131.216) (28.0,131.209) (29.0,131.216) (30.0,131.214) (31.0,131.215) (32.0,131.214) (33.0,131.215) };
\addlegendentry{B-tree BS, $B=8$};
\addplot coordinates { (27.0,163.119) (28.0,163.12) (29.0,163.119) (30.0,163.125) (31.0,163.122) (32.0,163.122) (33.0,163.123) };
\addlegendentry{Fusion Tree SIMD, $k=16$};
\addplot coordinates { (27.0,167.708) (28.0,167.699) (29.0,167.708) (30.0,167.705) (31.0,167.707) (32.0,167.705) (33.0,167.706) };
\addlegendentry{Fusion Tree SIMD, $k=8$};
\addplot coordinates { (27.0,163.119) (28.0,163.12) (29.0,163.119) (30.0,163.125) (31.0,163.122) (32.0,163.122) (33.0,163.123) };
\addlegendentry{Fusion Tree LS, $k=16$};
\addplot coordinates { (27.0,167.708) (28.0,167.699) (29.0,167.708) (30.0,167.705) (31.0,167.707) (32.0,167.705) (33.0,167.706) };
\addlegendentry{Fusion Tree LS, $k=8$};
\legend{}
\end{axis}
\end{tikzpicture}
\centering\ref{leg:fusion:u64}
\caption{
Throughputs for the insert, predecessor and delete operations, as well as memory usage of the fusion trees and B-trees for $U=[2^{64}]$.
Best viewed in colour.
In the legend, \emph{LS} stands for linear searched nodes, \emph{BS} for binary search and \emph{SIMD} for use of SIMD instructions.
}
\label{fig:fusion:u64}
\end{figure}
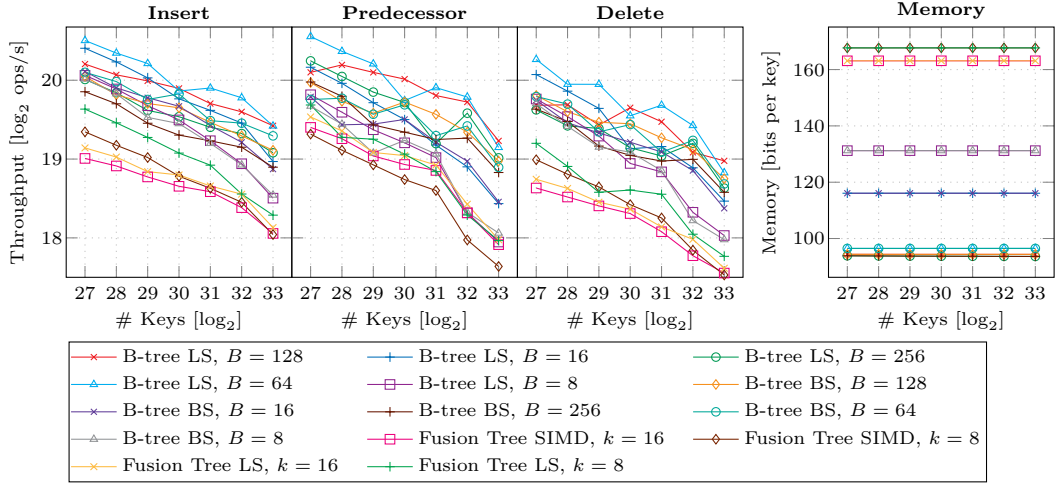

\subparagraph*{Experimental evaluation.}
We use $B:=8$ and $B:=16$ for fusion trees with $k=8$ and $k=16$, respectively, comparing fusion nodes using the \pcmpgtb{} instruction for rank queries against those using simple linear scans.
Further, we compare fusion trees to straight B-trees, finding predecessors in a node using $\Oh{\lg B}$-time binary or $\Oh{B}$-time linear search.
There, we also consider much larger $B$, as preliminary experiments suggested that the performance of all operations peaks at $B:=64$.
Our results for 64-bit keys are presented in \autoref{fig:fusion:u64}.
(More results for smaller universes are given in \autoref{fig:fusion:u40} in \autoref{appendix:results}.)

To our surprise, fusion trees achieve the lowest throughputs for all operations: B-trees with large $B$ are up to twice as fast, and even the B-trees with low degrees are visibly faster overall.
Fusion trees also require more memory per key, which was expected, as each node needs to store three words (the compression mask and two matrices) in addition to the keys themselves.
Interestingly, the fusion nodes using linear scans for ranking outperform those that use the SIMD instructions in nearly all instances.
The reason is presumably that the corresponding MMX/AVX registers have to be filled prior to executing these instructions: in a direct comparison answering immediately consecutive random rank queries, the SIMD variant is about 28\% faster than scanning.
Fusion trees with $B=16$ perform slower overall than those with $B=8$ despite their lower height, which is due to overheads in our simulation of 256-bit words.
It shall be interesting to redo these experiments with natively supported wide registers (e.g., AVX-512) and necessary instructions in the future.

We have a brief closer look at B-trees.
Our preliminary experiments are largely confirmed in that B-trees with $B=64$ perform best overall.
For $B \leq 64$, nodes backed by linear search perform faster than those backed by binary search.
The exact opposite is the case for $B>64$, where binary search becomes faster.
Concerning memory, unsurprisingly, the higher $B$ is chosen, the less memory is required as the tree structure shrinks in height.

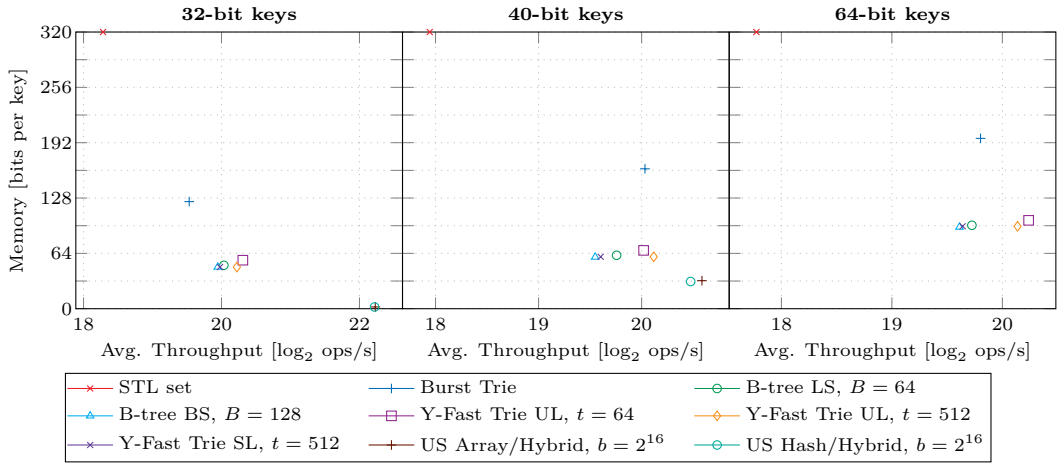
\begin{figure}
\begin{tikzpicture}
\begin{groupplot}[group3wide, ybitsperkey, ymin=0, ymax=320, ytick={0,32,64,96,128,160,192,224,256,288,320}, yticklabels={0,,64,,128,,192,,256,,320}]
\nextgroupplot[xavglogtput, title={32-bit keys}, legend to name=leg:comparison]
\addplot coordinates { (18.2827,320) };
\addlegendentry{STL~set};
\addplot coordinates { (19.5319,123.887) };
\addlegendentry{Burst Trie};
\addplot coordinates { (20.0344,50.1908) };
\addlegendentry{B-tree LS, $B=64$};
\addplot coordinates { (19.946,48.1471) };
\addlegendentry{B-tree BS, $B=128$};
\addplot coordinates { (20.3096,56.0884) };
\addlegendentry{Y-Fast Trie UL, $t=64$};
\addplot coordinates { (20.223,48.1648) };
\addlegendentry{Y-Fast Trie UL, $t=512$};
\addplot coordinates { (19.9784,48.1648) };
\addlegendentry{Y-Fast Trie SL, $t=512$};
\addplot coordinates { (22.2296,2.01167) };
\addlegendentry{US Array/Hybrid, $b=2^{16}$};
\addplot coordinates { (22.2193,2.00993) };
\addlegendentry{US Hash/Hybrid, $b=2^{16}$};
\nextgroupplot[xavglogtput, title={40-bit keys}]
\addplot coordinates { (17.9436,320) };
\addlegendentry{STL~set};
\addplot coordinates { (20.0325,161.796) };
\addlegendentry{Burst Trie};
\addplot coordinates { (19.7572,61.7754) };
\addlegendentry{B-tree LS, $B=64$};
\addplot coordinates { (19.549,59.7044) };
\addlegendentry{B-tree BS, $B=128$};
\addplot coordinates { (20.0193,67.4377) };
\addlegendentry{Y-Fast Trie UL, $t=64$};
\addplot coordinates { (20.1176,60.0471) };
\addlegendentry{Y-Fast Trie UL, $t=512$};
\addplot coordinates { (19.6025,60.0471) };
\addlegendentry{Y-Fast Trie SL, $t=512$};
\addplot coordinates { (20.5855,32.4174) };
\addlegendentry{US Array/Hybrid, $b=2^{16}$};
\addplot coordinates { (20.476,31.4445) };
\addlegendentry{US Hash/Hybrid, $b=2^{16}$};
\legend{}
\nextgroupplot[xavglogtput, title={64-bit keys}]
\addplot coordinates { (17.7739,320) };
\addlegendentry{STL~set};
\addplot coordinates { (19.8024,196.996) };
\addlegendentry{Burst Trie};
\addplot coordinates { (19.7241,96.4889) };
\addlegendentry{B-tree LS, $B=64$};
\addplot coordinates { (19.6118,94.3742) };
\addlegendentry{B-tree BS, $B=128$};
\addplot coordinates { (20.2377,102.193) };
\addlegendentry{Y-Fast Trie UL, $t=64$};
\addplot coordinates { (20.1374,95.3288) };
\addlegendentry{Y-Fast Trie UL, $t=512$};
\addplot coordinates { (19.6404,95.3288) };
\addlegendentry{Y-Fast Trie SL, $t=512$};
\legend{}
\end{groupplot}
\end{tikzpicture}
\centering\ref{leg:comparison}
\caption{Comparing the average throughput of operations versus memory use of dynamic predecessor data structures for different universes and $n=2^{30}$.}
\label{fig:comparison}
\end{figure}

\section{Comparison}
\label{sec:comparison}

In \autoref{fig:comparison}, we plot the average throughput of insertions, predecessor queries and deletions against the memory usage of a subset of our data structures from Sections~\ref{sec:sampling}--\ref{sec:fusion} for a fixed workload size of $2^{30}$.
For comparison, we also show the performance of the STL set (\texttt{std::set}, an implementation of red-black trees), and the burst trie of Nash and Gregg \cite{DBLP:journals/jea/NashG10}---to the best of our knowledge the best practical dynamic predecessor data structure thus far.
Note that burst tries are \emph{associative} and store a value along with each key, so for a fair comparison, one should subtract $w$ bits per key for each data point.

For all universes, at least one of our data structures is over four times faster than the STL set, the extreme being for 32-bit keys, where our sampling structures achieve an average throughput of approximately $2^{22.2}$ operations per second, whereas the set does about $2^{18.2}$.
Furthermore, our data structures consume less than a third of the set's 320 bits per key.
For 32-bit keys, we also outperform burst tries completely, where even our slowest data structure (B-trees with degree 128 and binary searched nodes) is about 33\% faster.
Our two sampling data structures with hybrid buckets of size $2^{16}$ are clearly the fastest.
Their low space consumption of just about 2 bits per key should, however, be interpreted with care, as for $n=2^{30}$, one quarter of all possible keys is contained in $S$, and hence they essentially store $S$ in a bit vector.
This also shows up for $w=40$ (but less pronounced), where they become the only data structure clearly faster than the burst tries.
Our y-fast tries with unsorted buckets of size $2^9$ are about 6\% faster than burst tries, but still require considerably less memory even respecting that 40 bits per key in burst tries are for associated values.
It appears that y-fast tries with unsorted buckets scale best with the size of the universe:
for 64-bit keys, it is the fastest data structure with buckets of size either $2^6$ or $2^9$ and is approximately 30\% faster than burst tries, again consuming significantly less memory.
The y-fast tries with sorted buckets are about on par with our B-trees, which are overall between 14\% and 56\% slower than y-fast tries with unsorted buckets.

For 32-bit keys, we intended to include the stratified tree \cite{DBLP:conf/alenex/DementievKMS04}, but it failed to stay within the memory limits (\SI{256}{\giga\byte}) starting at $2^{30}$ keys.
For $2^{29}$ keys, it consumed about \num{1480} bits per key, ranking lowest with an average throughput of circa $2^{17.7}$ operations per second.

\subparagraph*{Conclusions.}
Our dynamic predecessor data structures are the most memory efficient of all tested.
They clearly outperform the STL set and for all universes in question, at least one of our data structures is faster than burst tries, the previously fastest known to us.
We confirm once more \cite{DBLP:conf/wea/BaumstarkGHL17,DBLP:conf/esa/Dinklage0HKK20,DBLP:journals/jda/FerradaN17,DBLP:conf/spire/IlieT09} that na\"ive solutions can be more practical than sophisticated data structures on modern hardware and sufficiently small inputs.
We also observed that SIMD instructions, while faster than sequences of \emph{classic} (SISD) instructions when used in batches, may turn out less useful in more complex scenarios.

\def\UrlBreaks{\do\/\do-}
\bibliography{literature}

\clearpage
\appendix

\begin{figure}[p]
\hspace*{\fill}\textbf{32-bit keys}\hspace*{\fill}

\begin{tikzpicture}
\begin{groupplot}[group3, lesshigh, ylogtput, ymin=18, ymax=22.5]
\nextgroupplot[xlognum, title={Insert}, legend to name=leg:yfast:u40]
\addplot coordinates { (27.0,21.2672) (28.0,21.0772) (29.0,20.8148) (30.0,20.7276) };
\addlegendentry{Y-Fast Trie UL, $t=64$};
\addplot coordinates { (27.0,21.4827) (28.0,21.2727) (29.0,20.9086) (30.0,20.6716) };
\addlegendentry{Y-Fast Trie UL, $t=128$};
\addplot coordinates { (27.0,21.7713) (28.0,21.4862) (29.0,21.2505) (30.0,21.0291) };
\addlegendentry{Y-Fast Trie UL, $t=256$};
\addplot coordinates { (27.0,22.0625) (28.0,21.7699) (29.0,21.4509) (30.0,21.1267) };
\addlegendentry{Y-Fast Trie UL, $t=512$};
\addplot coordinates { (27.0,20.6562) (28.0,20.5231) (29.0,20.3829) (30.0,20.2209) };
\addlegendentry{Y-Fast Trie SL, $t=64$};
\addplot coordinates { (27.0,20.5206) (28.0,20.4113) (29.0,20.2891) (30.0,20.1619) };
\addlegendentry{Y-Fast Trie SL, $t=128$};
\addplot coordinates { (27.0,20.2912) (28.0,20.0327) (29.0,20.025) (30.0,19.9643) };
\addlegendentry{Y-Fast Trie SL, $t=256$};
\addplot coordinates { (27.0,19.9135) (28.0,19.7324) (29.0,19.6961) (30.0,19.6171) };
\addlegendentry{Y-Fast Trie SL, $t=512$};
\nextgroupplot[xlognum, title={Predecessor}]
\addplot coordinates { (27.0,20.1418) (28.0,20.0059) (29.0,19.8227) (30.0,19.814) };
\addlegendentry{Y-Fast Trie UL, $t=64$};
\addplot coordinates { (27.0,19.7222) (28.0,19.633) (29.0,19.3737) (30.0,19.3452) };
\addlegendentry{Y-Fast Trie UL, $t=128$};
\addplot coordinates { (27.0,19.146) (28.0,19.0957) (29.0,19.0338) (30.0,18.9655) };
\addlegendentry{Y-Fast Trie UL, $t=256$};
\addplot coordinates { (27.0,18.4416) (28.0,18.4178) (29.0,18.3963) (30.0,18.3007) };
\addlegendentry{Y-Fast Trie UL, $t=512$};
\addplot coordinates { (27.0,20.642) (28.0,20.4968) (29.0,20.3299) (30.0,20.0253) };
\addlegendentry{Y-Fast Trie SL, $t=64$};
\addplot coordinates { (27.0,20.6046) (28.0,20.2772) (29.0,20.3267) (30.0,20.1924) };
\addlegendentry{Y-Fast Trie SL, $t=128$};
\addplot coordinates { (27.0,20.5438) (28.0,20.2999) (29.0,20.2973) (30.0,20.1793) };
\addlegendentry{Y-Fast Trie SL, $t=256$};
\addplot coordinates { (27.0,20.5473) (28.0,20.1633) (29.0,20.021) (30.0,20.1914) };
\addlegendentry{Y-Fast Trie SL, $t=512$};
\legend{}
\nextgroupplot[xlognum, title={Delete}]
\addplot coordinates { (27.0,20.7308) (28.0,20.5439) (29.0,20.2868) (30.0,20.2427) };
\addlegendentry{Y-Fast Trie UL, $t=64$};
\addplot coordinates { (27.0,20.7482) (28.0,20.4447) (29.0,20.2017) (30.0,19.9368) };
\addlegendentry{Y-Fast Trie UL, $t=128$};
\addplot coordinates { (27.0,20.6612) (28.0,20.5163) (29.0,20.3665) (30.0,20.1991) };
\addlegendentry{Y-Fast Trie UL, $t=256$};
\addplot coordinates { (27.0,20.4897) (28.0,20.3151) (29.0,20.2446) (30.0,20.0142) };
\addlegendentry{Y-Fast Trie UL, $t=512$};
\addplot coordinates { (27.0,20.6367) (28.0,20.4444) (29.0,20.3014) (30.0,20.0048) };
\addlegendentry{Y-Fast Trie SL, $t=64$};
\addplot coordinates { (27.0,20.6282) (28.0,20.2961) (29.0,20.3194) (30.0,20.1885) };
\addlegendentry{Y-Fast Trie SL, $t=128$};
\addplot coordinates { (27.0,20.5522) (28.0,20.3182) (29.0,20.2784) (30.0,20.1547) };
\addlegendentry{Y-Fast Trie SL, $t=256$};
\addplot coordinates { (27.0,20.4114) (28.0,19.9484) (29.0,19.9827) (30.0,20.0659) };
\addlegendentry{Y-Fast Trie SL, $t=512$};
\legend{}
\end{groupplot}
\end{tikzpicture}
\begin{tikzpicture}
\begin{axis}[plot4col, lesshigh, xlognum, ybitsperkey, title={Memory}]
\addplot coordinates { (27.0,56.1258) (28.0,56.1118) (29.0,56.1087) (30.0,56.0884) };
\addlegendentry{Y-Fast Trie UL, $t=64$};
\addplot coordinates { (27.0,51.3363) (28.0,51.3063) (29.0,51.2868) (30.0,51.2616) };
\addlegendentry{Y-Fast Trie UL, $t=128$};
\addplot coordinates { (27.0,49.2623) (28.0,49.2332) (29.0,49.2037) (30.0,49.1753) };
\addlegendentry{Y-Fast Trie UL, $t=256$};
\addplot coordinates { (27.0,48.1915) (28.0,48.1859) (29.0,48.1781) (30.0,48.1648) };
\addlegendentry{Y-Fast Trie UL, $t=512$};
\addplot coordinates { (27.0,56.1258) (28.0,56.1118) (29.0,56.1087) (30.0,56.0884) };
\addlegendentry{Y-Fast Trie SL, $t=64$};
\addplot coordinates { (27.0,51.3363) (28.0,51.3063) (29.0,51.2868) (30.0,51.2616) };
\addlegendentry{Y-Fast Trie SL, $t=128$};
\addplot coordinates { (27.0,49.2623) (28.0,49.2332) (29.0,49.2037) (30.0,49.1753) };
\addlegendentry{Y-Fast Trie SL, $t=256$};
\addplot coordinates { (27.0,48.1915) (28.0,48.1859) (29.0,48.1781) (30.0,48.1648) };
\addlegendentry{Y-Fast Trie SL, $t=512$};
\legend{}
\end{axis}
\end{tikzpicture}

\hspace*{\fill}\textbf{40-bit keys}\hspace*{\fill}

\begin{tikzpicture}
\begin{groupplot}[group3, lesshigh, ylogtput, ymin=17, ymax=22]
\nextgroupplot[xlognum, title={Insert}]
\addplot coordinates { (27.0,21.0848) (28.0,20.8878) (29.0,20.7978) (30.0,20.5086) (31.0,20.4636) (32.0,20.2562) (33.0,20.0726) };
\addlegendentry{Y-Fast Trie UL, $t=64$};
\addplot coordinates { (27.0,21.2759) (28.0,21.1025) (29.0,20.9196) (30.0,20.6846) (31.0,20.4579) (32.0,20.3289) (33.0,20.2433) };
\addlegendentry{Y-Fast Trie UL, $t=128$};
\addplot coordinates { (27.0,21.5316) (28.0,21.2933) (29.0,20.9849) (30.0,20.7563) (31.0,20.7291) (32.0,20.4703) (33.0,20.2448) };
\addlegendentry{Y-Fast Trie UL, $t=256$};
\addplot coordinates { (27.0,21.8058) (28.0,21.5083) (29.0,21.253) (30.0,21.0262) (31.0,20.8206) (32.0,20.7055) (33.0,20.3924) };
\addlegendentry{Y-Fast Trie UL, $t=512$};
\addplot coordinates { (27.0,20.4568) (28.0,20.3574) (29.0,20.2309) (30.0,20.1346) (31.0,20.0251) (32.0,19.5339) (33.0,19.6245) };
\addlegendentry{Y-Fast Trie SL, $t=64$};
\addplot coordinates { (27.0,20.2713) (28.0,20.1656) (29.0,19.9828) (30.0,19.9345) (31.0,19.8798) (32.0,19.7864) (33.0,19.4365) };
\addlegendentry{Y-Fast Trie SL, $t=128$};
\addplot coordinates { (27.0,19.9355) (28.0,19.7288) (29.0,19.7524) (30.0,19.669) (31.0,19.538) (32.0,19.307) (33.0,19.2277) };
\addlegendentry{Y-Fast Trie SL, $t=256$};
\addplot coordinates { (27.0,19.4162) (28.0,19.2719) (29.0,19.22) (30.0,19.0898) (31.0,19.0109) (32.0,19.0496) (33.0,18.8696) };
\addlegendentry{Y-Fast Trie SL, $t=512$};
\legend{}
\nextgroupplot[xlognum, title={Predecessor}]
\addplot coordinates { (27.0,19.7596) (28.0,19.6417) (29.0,19.5796) (30.0,19.1987) (31.0,19.447) (32.0,19.384) (33.0,19.0834) };
\addlegendentry{Y-Fast Trie UL, $t=64$};
\addplot coordinates { (27.0,19.2411) (28.0,19.1821) (29.0,19.1226) (30.0,19.0312) (31.0,18.8005) (32.0,18.9396) (33.0,18.6248) };
\addlegendentry{Y-Fast Trie UL, $t=128$};
\addplot coordinates { (27.0,18.4477) (28.0,18.4829) (29.0,18.4395) (30.0,18.3656) (31.0,18.3262) (32.0,18.3922) (33.0,18.0788) };
\addlegendentry{Y-Fast Trie UL, $t=256$};
\addplot coordinates { (27.0,17.6272) (28.0,17.5913) (29.0,17.585) (30.0,17.6309) (31.0,17.6901) (32.0,17.6715) (33.0,17.5095) };
\addlegendentry{Y-Fast Trie UL, $t=512$};
\addplot coordinates { (27.0,20.5196) (28.0,20.4418) (29.0,20.2192) (30.0,20.1483) (31.0,20.0366) (32.0,19.6149) (33.0,19.4633) };
\addlegendentry{Y-Fast Trie SL, $t=64$};
\addplot coordinates { (27.0,20.5087) (28.0,20.3771) (29.0,20.1126) (30.0,20.0602) (31.0,20.0476) (32.0,19.9526) (33.0,19.4993) };
\addlegendentry{Y-Fast Trie SL, $t=128$};
\addplot coordinates { (27.0,20.4333) (28.0,20.1299) (29.0,20.2257) (30.0,20.1359) (31.0,19.7944) (32.0,19.7215) (33.0,19.5619) };
\addlegendentry{Y-Fast Trie SL, $t=256$};
\addplot coordinates { (27.0,20.1978) (28.0,20.1413) (29.0,20.1422) (30.0,19.8821) (31.0,19.8324) (32.0,19.5203) (33.0,19.5103) };
\addlegendentry{Y-Fast Trie SL, $t=512$};
\legend{}
\nextgroupplot[xlognum, title={Delete}]
\addplot coordinates { (27.0,20.6162) (28.0,20.4177) (29.0,20.3067) (30.0,20.0621) (31.0,19.8449) (32.0,19.9777) (33.0,19.5229) };
\addlegendentry{Y-Fast Trie UL, $t=64$};
\addplot coordinates { (27.0,20.6166) (28.0,20.4749) (29.0,20.3219) (30.0,19.8773) (31.0,19.9189) (32.0,19.9049) (33.0,19.2921) };
\addlegendentry{Y-Fast Trie UL, $t=128$};
\addplot coordinates { (27.0,20.5281) (28.0,20.3405) (29.0,20.1345) (30.0,20.0272) (31.0,20.023) (32.0,19.9069) (33.0,19.164) };
\addlegendentry{Y-Fast Trie UL, $t=256$};
\addplot coordinates { (27.0,20.0669) (28.0,20.2233) (29.0,20.1652) (30.0,20.035) (31.0,19.9403) (32.0,19.5825) (33.0,19.348) };
\addlegendentry{Y-Fast Trie UL, $t=512$};
\addplot coordinates { (27.0,20.5105) (28.0,20.3954) (29.0,20.197) (30.0,20.1242) (31.0,19.8956) (32.0,19.5983) (33.0,19.4015) };
\addlegendentry{Y-Fast Trie SL, $t=64$};
\addplot coordinates { (27.0,20.5154) (28.0,20.3718) (29.0,20.1129) (30.0,20.0642) (31.0,19.857) (32.0,19.7064) (33.0,19.4676) };
\addlegendentry{Y-Fast Trie SL, $t=128$};
\addplot coordinates { (27.0,20.3812) (28.0,20.0305) (29.0,20.1547) (30.0,20.0639) (31.0,19.7728) (32.0,19.6728) (33.0,19.5105) };
\addlegendentry{Y-Fast Trie SL, $t=256$};
\addplot coordinates { (27.0,20.0402) (28.0,19.99) (29.0,19.9395) (30.0,19.7204) (31.0,19.6475) (32.0,19.2586) (33.0,19.3414) };
\addlegendentry{Y-Fast Trie SL, $t=512$};
\legend{}
\end{groupplot}
\end{tikzpicture}
\begin{tikzpicture}
\begin{axis}[plot4col, lesshigh, xlognum, ybitsperkey, title={Memory}]
\addplot coordinates { (27.0,67.441) (28.0,67.443) (29.0,67.4389) (30.0,67.4377) (31.0,67.4375) (32.0,67.4382) (33.0,67.4384) };
\addlegendentry{Y-Fast Trie UL, $t=64$};
\addplot coordinates { (27.0,62.7226) (28.0,62.7115) (29.0,62.6991) (30.0,62.6818) (31.0,62.668) (32.0,62.6586) (33.0,62.6486) };
\addlegendentry{Y-Fast Trie UL, $t=128$};
\addplot coordinates { (27.0,60.7701) (28.0,60.7416) (29.0,60.7084) (30.0,60.6821) (31.0,60.6583) (32.0,60.6362) (33.0,60.6126) };
\addlegendentry{Y-Fast Trie UL, $t=256$};
\addplot coordinates { (27.0,60.1515) (28.0,60.1159) (29.0,60.075) (30.0,60.0471) (31.0,60.0235) (32.0,59.9925) (33.0,59.9667) };
\addlegendentry{Y-Fast Trie UL, $t=512$};
\addplot coordinates { (27.0,67.441) (28.0,67.443) (29.0,67.4389) (30.0,67.4377) (31.0,67.4375) (32.0,67.4382) (33.0,67.4384) };
\addlegendentry{Y-Fast Trie SL, $t=64$};
\addplot coordinates { (27.0,62.7226) (28.0,62.7115) (29.0,62.6991) (30.0,62.6818) (31.0,62.668) (32.0,62.6586) (33.0,62.6486) };
\addlegendentry{Y-Fast Trie SL, $t=128$};
\addplot coordinates { (27.0,60.7701) (28.0,60.7416) (29.0,60.7084) (30.0,60.6821) (31.0,60.6583) (32.0,60.6362) (33.0,60.6126) };
\addlegendentry{Y-Fast Trie SL, $t=256$};
\addplot coordinates { (27.0,60.1515) (28.0,60.1159) (29.0,60.075) (30.0,60.0471) (31.0,60.0235) (32.0,59.9925) (33.0,59.9667) };
\addlegendentry{Y-Fast Trie SL, $t=512$};
\legend{}
\end{axis}
\end{tikzpicture}
\centering\ref{leg:yfast:u40}
\caption{Throughputs for the insert, predecessor and delete operations, as well as memory usage of the y-fast trie for 32-bit (top) and 40-bit keys (bottom). Best viewed in colour.}
\label{fig:yfast:u40}
\end{figure}
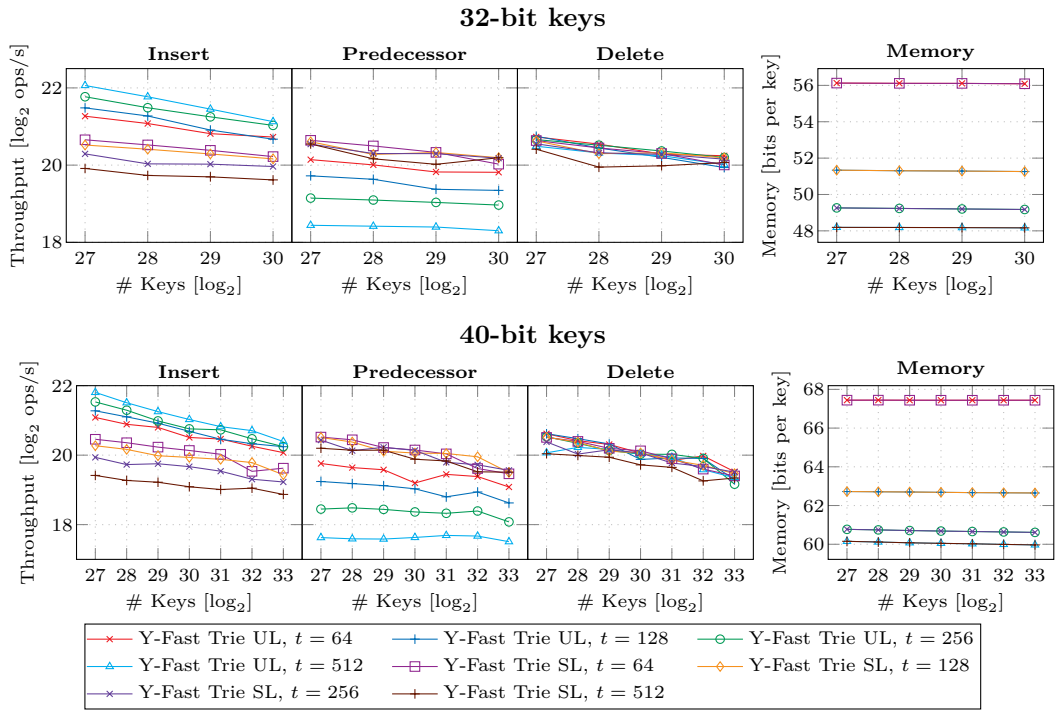

\begin{figure}[p]
\hspace*{\fill}\textbf{32-bit keys}\hspace*{\fill}

\begin{tikzpicture}
\begin{groupplot}[group3, lesshigh, ylogtput, ymin=18, ymax=21]
\nextgroupplot[xlognum, title={Insert}, legend to name=leg:fusion:u40]
\addplot coordinates { (27.0,20.662) (28.0,20.5062) (29.0,20.3516) (30.0,20.1661) };
\addlegendentry{B-tree LS, $B=128$};
\addplot coordinates { (27.0,20.5534) (28.0,20.4002) (29.0,20.2184) (30.0,19.9234) };
\addlegendentry{B-tree LS, $B=16$};
\addplot coordinates { (27.0,20.3207) (28.0,20.1894) (29.0,19.9674) (30.0,19.9611) };
\addlegendentry{B-tree LS, $B=256$};
\addplot coordinates { (27.0,20.7816) (28.0,20.5874) (29.0,20.4845) (30.0,20.2579) };
\addlegendentry{B-tree LS, $B=64$};
\addplot coordinates { (27.0,20.1917) (28.0,20.0393) (29.0,19.822) (30.0,19.5555) };
\addlegendentry{B-tree LS, $B=8$};
\addplot coordinates { (27.0,20.4692) (28.0,20.3212) (29.0,20.1098) (30.0,19.9763) };
\addlegendentry{B-tree BS, $B=128$};
\addplot coordinates { (27.0,20.4325) (28.0,20.2692) (29.0,20.1086) (30.0,19.9049) };
\addlegendentry{B-tree BS, $B=16$};
\addplot coordinates { (27.0,20.1413) (28.0,20.0146) (29.0,19.8942) (30.0,19.8023) };
\addlegendentry{B-tree BS, $B=256$};
\addplot coordinates { (27.0,20.571) (28.0,20.4462) (29.0,20.3304) (30.0,20.1617) };
\addlegendentry{B-tree BS, $B=64$};
\addplot coordinates { (27.0,20.1308) (28.0,19.9951) (29.0,19.7658) (30.0,19.5674) };
\addlegendentry{B-tree BS, $B=8$};
\addplot coordinates { (27.0,19.0701) (28.0,18.9796) (29.0,18.8686) (30.0,18.7554) };
\addlegendentry{Fusion Tree SIMD, $k=16$};
\addplot coordinates { (27.0,19.4026) (28.0,19.1471) (29.0,19.1079) (30.0,18.9528) };
\addlegendentry{Fusion Tree SIMD, $k=8$};
\addplot coordinates { (27.0,19.2093) (28.0,19.1046) (29.0,18.9994) (30.0,18.8705) };
\addlegendentry{Fusion Tree LS, $k=16$};
\addplot coordinates { (27.0,19.6758) (28.0,19.5964) (29.0,19.4139) (30.0,19.1649) };
\addlegendentry{Fusion Tree LS, $k=8$};
\nextgroupplot[xlognum, title={Predecessor}]
\addplot coordinates { (27.0,20.771) (28.0,20.583) (29.0,20.3853) (30.0,20.2059) };
\addlegendentry{B-tree LS, $B=128$};
\addplot coordinates { (27.0,20.3681) (28.0,20.2121) (29.0,19.9744) (30.0,19.5723) };
\addlegendentry{B-tree LS, $B=16$};
\addplot coordinates { (27.0,20.618) (28.0,20.4627) (29.0,20.1952) (30.0,20.0775) };
\addlegendentry{B-tree LS, $B=256$};
\addplot coordinates { (27.0,20.8036) (28.0,20.6024) (29.0,20.4675) (30.0,20.0333) };
\addlegendentry{B-tree LS, $B=64$};
\addplot coordinates { (27.0,20.041) (28.0,19.8513) (29.0,19.5916) (30.0,19.1763) };
\addlegendentry{B-tree LS, $B=8$};
\addplot coordinates { (27.0,20.6163) (28.0,20.461) (29.0,20.2093) (30.0,20.1189) };
\addlegendentry{B-tree BS, $B=128$};
\addplot coordinates { (27.0,20.2971) (28.0,20.1343) (29.0,19.9195) (30.0,19.6882) };
\addlegendentry{B-tree BS, $B=16$};
\addplot coordinates { (27.0,20.4151) (28.0,20.2534) (29.0,20.0838) (30.0,19.9513) };
\addlegendentry{B-tree BS, $B=256$};
\addplot coordinates { (27.0,20.476) (28.0,20.4436) (29.0,20.2945) (30.0,20.1195) };
\addlegendentry{B-tree BS, $B=64$};
\addplot coordinates { (27.0,19.9841) (28.0,19.8086) (29.0,19.5466) (30.0,19.3603) };
\addlegendentry{B-tree BS, $B=8$};
\addplot coordinates { (27.0,19.4875) (28.0,19.3399) (29.0,19.1404) (30.0,19.0178) };
\addlegendentry{Fusion Tree SIMD, $k=16$};
\addplot coordinates { (27.0,19.4269) (28.0,18.791) (29.0,19.037) (30.0,18.9053) };
\addlegendentry{Fusion Tree SIMD, $k=8$};
\addplot coordinates { (27.0,19.646) (28.0,19.4825) (29.0,19.3024) (30.0,18.8989) };
\addlegendentry{Fusion Tree LS, $k=16$};
\addplot coordinates { (27.0,19.7438) (28.0,19.6495) (29.0,19.412) (30.0,19.1896) };
\addlegendentry{Fusion Tree LS, $k=8$};
\legend{}
\nextgroupplot[xlognum, title={Delete}]
\addplot coordinates { (27.0,20.3247) (28.0,20.1559) (29.0,20.0001) (30.0,19.916) };
\addlegendentry{B-tree LS, $B=128$};
\addplot coordinates { (27.0,20.2693) (28.0,20.096) (29.0,19.8575) (30.0,19.4913) };
\addlegendentry{B-tree LS, $B=16$};
\addplot coordinates { (27.0,20.0753) (28.0,19.9371) (29.0,19.7366) (30.0,19.6603) };
\addlegendentry{B-tree LS, $B=256$};
\addplot coordinates { (27.0,20.4708) (28.0,20.2755) (29.0,20.1325) (30.0,19.7712) };
\addlegendentry{B-tree LS, $B=64$};
\addplot coordinates { (27.0,19.91) (28.0,19.706) (29.0,19.4623) (30.0,18.8926) };
\addlegendentry{B-tree LS, $B=8$};
\addplot coordinates { (27.0,20.2378) (28.0,20.0861) (29.0,19.8769) (30.0,19.7139) };
\addlegendentry{B-tree BS, $B=128$};
\addplot coordinates { (27.0,20.0868) (28.0,19.9775) (29.0,19.7692) (30.0,19.568) };
\addlegendentry{B-tree BS, $B=16$};
\addplot coordinates { (27.0,19.9498) (28.0,19.803) (29.0,19.6662) (30.0,19.5565) };
\addlegendentry{B-tree BS, $B=256$};
\addplot coordinates { (27.0,20.252) (28.0,20.1703) (29.0,20.0202) (30.0,19.8536) };
\addlegendentry{B-tree BS, $B=64$};
\addplot coordinates { (27.0,19.8005) (28.0,19.4676) (29.0,19.3573) (30.0,18.9292) };
\addlegendentry{B-tree BS, $B=8$};
\addplot coordinates { (27.0,18.6984) (28.0,18.5939) (29.0,18.4339) (30.0,18.3787) };
\addlegendentry{Fusion Tree SIMD, $k=16$};
\addplot coordinates { (27.0,19.0448) (28.0,18.4874) (29.0,18.3785) (30.0,18.5839) };
\addlegendentry{Fusion Tree SIMD, $k=8$};
\addplot coordinates { (27.0,18.8101) (28.0,18.6918) (29.0,18.5791) (30.0,18.3073) };
\addlegendentry{Fusion Tree LS, $k=16$};
\addplot coordinates { (27.0,19.1973) (28.0,19.1938) (29.0,19.0036) (30.0,18.7905) };
\addlegendentry{Fusion Tree LS, $k=8$};
\legend{}
\end{groupplot}
\end{tikzpicture}
\begin{tikzpicture}
\begin{axis}[plot4col, lesshigh, xlognum, ybitsperkey, title={Memory}]
\addplot coordinates { (27.0,48.1462) (28.0,48.1466) (29.0,48.1502) (30.0,48.1471) };
\addlegendentry{B-tree LS, $B=128$};
\addplot coordinates { (27.0,66.1074) (28.0,66.1038) (29.0,66.1033) (30.0,66.1062) };
\addlegendentry{B-tree LS, $B=16$};
\addplot coordinates { (27.0,47.4408) (28.0,47.4168) (29.0,47.3931) (30.0,47.3788) };
\addlegendentry{B-tree LS, $B=256$};
\addplot coordinates { (27.0,50.1891) (28.0,50.1869) (29.0,50.1907) (30.0,50.1908) };
\addlegendentry{B-tree LS, $B=64$};
\addplot coordinates { (27.0,82.5556) (28.0,82.5569) (29.0,82.5592) (30.0,82.5578) };
\addlegendentry{B-tree LS, $B=8$};
\addplot coordinates { (27.0,48.1462) (28.0,48.1466) (29.0,48.1502) (30.0,48.1471) };
\addlegendentry{B-tree BS, $B=128$};
\addplot coordinates { (27.0,66.1074) (28.0,66.1038) (29.0,66.1033) (30.0,66.1062) };
\addlegendentry{B-tree BS, $B=16$};
\addplot coordinates { (27.0,47.4408) (28.0,47.4168) (29.0,47.3931) (30.0,47.3788) };
\addlegendentry{B-tree BS, $B=256$};
\addplot coordinates { (27.0,50.1891) (28.0,50.1869) (29.0,50.1907) (30.0,50.1908) };
\addlegendentry{B-tree BS, $B=64$};
\addplot coordinates { (27.0,82.5556) (28.0,82.5569) (29.0,82.5592) (30.0,82.5578) };
\addlegendentry{B-tree BS, $B=8$};
\addplot coordinates { (27.0,113.149) (28.0,113.143) (29.0,113.142) (30.0,113.147) };
\addlegendentry{Fusion Tree SIMD, $k=16$};
\addplot coordinates { (27.0,112.964) (28.0,112.966) (29.0,112.969) (30.0,112.967) };
\addlegendentry{Fusion Tree SIMD, $k=8$};
\addplot coordinates { (27.0,113.149) (28.0,113.143) (29.0,113.142) (30.0,113.147) };
\addlegendentry{Fusion Tree LS, $k=16$};
\addplot coordinates { (27.0,112.964) (28.0,112.966) (29.0,112.969) (30.0,112.967) };
\addlegendentry{Fusion Tree LS, $k=8$};
\legend{}
\end{axis}
\end{tikzpicture}

\hspace*{\fill}\textbf{40-bit keys}\hspace*{\fill}

\begin{tikzpicture}
\begin{groupplot}[group3, lesshigh, ylogtput, ymin=17.25, ymax=20.5]
\nextgroupplot[xlognum, title={Insert}]
\addplot coordinates { (27.0,20.2627) (28.0,20.1243) (29.0,19.8907) (30.0,19.8332) (31.0,19.6715) (32.0,19.6204) (33.0,19.3146) };
\addlegendentry{B-tree LS, $B=128$};
\addplot coordinates { (27.0,20.4023) (28.0,20.2524) (29.0,20.0729) (30.0,19.7777) (31.0,19.6172) (32.0,19.2925) (33.0,19.2017) };
\addlegendentry{B-tree LS, $B=16$};
\addplot coordinates { (27.0,19.9403) (28.0,19.8275) (29.0,19.6806) (30.0,19.4881) (31.0,19.4259) (32.0,19.2432) (33.0,19.1448) };
\addlegendentry{B-tree LS, $B=256$};
\addplot coordinates { (27.0,20.4596) (28.0,20.3112) (29.0,20.1456) (30.0,19.865) (31.0,19.8087) (32.0,19.7466) (33.0,19.3831) };
\addlegendentry{B-tree LS, $B=64$};
\addplot coordinates { (27.0,20.0973) (28.0,19.8903) (29.0,19.7373) (30.0,19.3217) (31.0,19.2192) (32.0,18.9447) (33.0,18.6582) };
\addlegendentry{B-tree LS, $B=8$};
\addplot coordinates { (27.0,20.1567) (28.0,19.9774) (29.0,19.7984) (30.0,19.6411) (31.0,19.5587) (32.0,19.2279) (33.0,19.2371) };
\addlegendentry{B-tree BS, $B=128$};
\addplot coordinates { (27.0,20.0753) (28.0,20.014) (29.0,19.7725) (30.0,19.5617) (31.0,19.4971) (32.0,19.2505) (33.0,18.8958) };
\addlegendentry{B-tree BS, $B=16$};
\addplot coordinates { (27.0,19.9087) (28.0,19.779) (29.0,19.5362) (30.0,19.5149) (31.0,19.1347) (32.0,19.2542) (33.0,18.9758) };
\addlegendentry{B-tree BS, $B=256$};
\addplot coordinates { (27.0,20.2922) (28.0,19.9702) (29.0,19.9894) (30.0,19.7336) (31.0,19.5793) (32.0,19.5451) (33.0,19.0123) };
\addlegendentry{B-tree BS, $B=64$};
\addplot coordinates { (27.0,20.0031) (28.0,19.8427) (29.0,19.5978) (30.0,19.3818) (31.0,19.312) (32.0,19.1293) (33.0,18.5552) };
\addlegendentry{B-tree BS, $B=8$};
\addplot coordinates { (27.0,18.8247) (28.0,18.7423) (29.0,18.5887) (30.0,18.4399) (31.0,18.3709) (32.0,18.3074) (33.0,18.0094) };
\addlegendentry{Fusion Tree SIMD, $k=16$};
\addplot coordinates { (27.0,19.2387) (28.0,19.1297) (29.0,18.8669) (30.0,18.8348) (31.0,18.6363) (32.0,18.5494) (33.0,18.1446) };
\addlegendentry{Fusion Tree SIMD, $k=8$};
\addplot coordinates { (27.0,18.9557) (28.0,18.8634) (29.0,18.7582) (30.0,18.5619) (31.0,18.5582) (32.0,18.4667) (33.0,18.1197) };
\addlegendentry{Fusion Tree LS, $k=16$};
\addplot coordinates { (27.0,19.3675) (28.0,19.2789) (29.0,19.1198) (30.0,18.9086) (31.0,18.69) (32.0,18.6857) (33.0,18.3165) };
\addlegendentry{Fusion Tree LS, $k=8$};
\legend{}
\nextgroupplot[xlognum, title={Predecessor}]
\addplot coordinates { (27.0,20.2866) (28.0,20.1077) (29.0,19.8039) (30.0,19.781) (31.0,19.6666) (32.0,19.576) (33.0,18.8594) };
\addlegendentry{B-tree LS, $B=128$};
\addplot coordinates { (27.0,20.2559) (28.0,20.0792) (29.0,19.832) (30.0,19.3135) (31.0,19.4546) (32.0,18.9243) (33.0,18.5468) };
\addlegendentry{B-tree LS, $B=16$};
\addplot coordinates { (27.0,20.0891) (28.0,19.9498) (29.0,19.7446) (30.0,19.5132) (31.0,19.4658) (32.0,19.1971) (33.0,19.0669) };
\addlegendentry{B-tree LS, $B=256$};
\addplot coordinates { (27.0,20.3933) (28.0,20.2186) (29.0,20.0607) (30.0,19.8069) (31.0,19.5779) (32.0,19.6545) (33.0,18.9277) };
\addlegendentry{B-tree LS, $B=64$};
\addplot coordinates { (27.0,19.9642) (28.0,19.5814) (29.0,19.5174) (30.0,18.8682) (31.0,18.7607) (32.0,18.7396) (33.0,18.1725) };
\addlegendentry{B-tree LS, $B=8$};
\addplot coordinates { (27.0,20.1532) (28.0,19.9415) (29.0,19.7396) (30.0,19.5967) (31.0,19.5128) (32.0,19.2747) (33.0,19.1555) };
\addlegendentry{B-tree BS, $B=128$};
\addplot coordinates { (27.0,19.8244) (28.0,19.8068) (29.0,19.528) (30.0,19.122) (31.0,19.2596) (32.0,18.999) (33.0,18.3961) };
\addlegendentry{B-tree BS, $B=16$};
\addplot coordinates { (27.0,20.1164) (28.0,19.9777) (29.0,19.7098) (30.0,19.6363) (31.0,19.2187) (32.0,19.3467) (33.0,19.0097) };
\addlegendentry{B-tree BS, $B=256$};
\addplot coordinates { (27.0,20.136) (28.0,19.8172) (29.0,19.8242) (30.0,19.5122) (31.0,19.4135) (32.0,19.3936) (33.0,18.7041) };
\addlegendentry{B-tree BS, $B=64$};
\addplot coordinates { (27.0,19.8646) (28.0,19.6548) (29.0,19.375) (30.0,19.1852) (31.0,19.0936) (32.0,18.6668) (33.0,18.0277) };
\addlegendentry{B-tree BS, $B=8$};
\addplot coordinates { (27.0,19.1857) (28.0,19.0517) (29.0,18.7182) (30.0,18.6116) (31.0,18.5608) (32.0,18.2551) (33.0,17.8481) };
\addlegendentry{Fusion Tree SIMD, $k=16$};
\addplot coordinates { (27.0,19.3418) (28.0,19.1902) (29.0,18.905) (30.0,18.8445) (31.0,18.655) (32.0,18.2909) (33.0,17.8623) };
\addlegendentry{Fusion Tree SIMD, $k=8$};
\addplot coordinates { (27.0,19.3626) (28.0,19.2445) (29.0,19.0506) (30.0,18.7636) (31.0,18.843) (32.0,18.5181) (33.0,18.0174) };
\addlegendentry{Fusion Tree LS, $k=16$};
\addplot coordinates { (27.0,19.4694) (28.0,19.2371) (29.0,19.1827) (30.0,18.9507) (31.0,18.6724) (32.0,18.3871) (33.0,18.0663) };
\addlegendentry{Fusion Tree LS, $k=8$};
\legend{}
\nextgroupplot[xlognum, title={Delete}]
\addplot coordinates { (27.0,19.9264) (28.0,19.7791) (29.0,19.5481) (30.0,19.5163) (31.0,19.4106) (32.0,19.2857) (33.0,18.6898) };
\addlegendentry{B-tree LS, $B=128$};
\addplot coordinates { (27.0,20.1228) (28.0,19.9398) (29.0,19.7178) (30.0,19.1044) (31.0,19.3738) (32.0,18.8709) (33.0,18.4509) };
\addlegendentry{B-tree LS, $B=16$};
\addplot coordinates { (27.0,19.4707) (28.0,19.4674) (29.0,19.3365) (30.0,19.1582) (31.0,18.9759) (32.0,18.7639) (33.0,18.5636) };
\addlegendentry{B-tree LS, $B=256$};
\addplot coordinates { (27.0,20.1262) (28.0,19.9605) (29.0,19.8023) (30.0,19.5847) (31.0,19.3861) (32.0,19.3401) (33.0,18.8005) };
\addlegendentry{B-tree LS, $B=64$};
\addplot coordinates { (27.0,19.8069) (28.0,19.4422) (29.0,19.1822) (30.0,18.764) (31.0,18.6676) (32.0,18.644) (33.0,18.0978) };
\addlegendentry{B-tree LS, $B=8$};
\addplot coordinates { (27.0,19.8623) (28.0,19.6789) (29.0,19.5269) (30.0,19.3977) (31.0,19.281) (32.0,19.0961) (33.0,18.7909) };
\addlegendentry{B-tree BS, $B=128$};
\addplot coordinates { (27.0,19.7291) (28.0,19.7146) (29.0,19.4588) (30.0,19.0915) (31.0,19.2079) (32.0,18.9578) (33.0,18.3625) };
\addlegendentry{B-tree BS, $B=16$};
\addplot coordinates { (27.0,19.5614) (28.0,19.4358) (29.0,19.2435) (30.0,19.0648) (31.0,18.8992) (32.0,19.0016) (33.0,18.7181) };
\addlegendentry{B-tree BS, $B=256$};
\addplot coordinates { (27.0,19.98) (28.0,19.7225) (29.0,19.6798) (30.0,19.0506) (31.0,19.234) (32.0,19.2873) (33.0,18.6124) };
\addlegendentry{B-tree BS, $B=64$};
\addplot coordinates { (27.0,19.6395) (28.0,19.424) (29.0,19.1705) (30.0,18.61) (31.0,18.9176) (32.0,18.2675) (33.0,17.8995) };
\addlegendentry{B-tree BS, $B=8$};
\addplot coordinates { (27.0,18.462) (28.0,18.3701) (29.0,18.1594) (30.0,18.0669) (31.0,17.9748) (32.0,17.7156) (33.0,17.5128) };
\addlegendentry{Fusion Tree SIMD, $k=16$};
\addplot coordinates { (27.0,18.8609) (28.0,18.7384) (29.0,18.5011) (30.0,18.4549) (31.0,18.2896) (32.0,17.9788) (33.0,17.5506) };
\addlegendentry{Fusion Tree SIMD, $k=8$};
\addplot coordinates { (27.0,18.542) (28.0,18.4443) (29.0,18.3288) (30.0,17.9549) (31.0,18.0582) (32.0,17.9177) (33.0,17.5649) };
\addlegendentry{Fusion Tree LS, $k=16$};
\addplot coordinates { (27.0,18.8369) (28.0,18.6952) (29.0,18.68) (30.0,18.4915) (31.0,18.2457) (32.0,18.0182) (33.0,17.7356) };
\addlegendentry{Fusion Tree LS, $k=8$};
\legend{}
\end{groupplot}
\end{tikzpicture}
\begin{tikzpicture}
\begin{axis}[plot4col, lesshigh, xlognum, ybitsperkey, title={Memory}]
\addplot coordinates { (27.0,59.7082) (28.0,59.7051) (29.0,59.7061) (30.0,59.7044) (31.0,59.7011) (32.0,59.7015) (33.0,59.7019) };
\addlegendentry{B-tree LS, $B=128$};
\addplot coordinates { (27.0,78.5999) (28.0,78.6022) (29.0,78.6019) (30.0,78.6003) (31.0,78.6003) (32.0,78.5985) (33.0,78.5991) };
\addlegendentry{B-tree LS, $B=16$};
\addplot coordinates { (27.0,58.9917) (28.0,58.9611) (29.0,58.9401) (30.0,58.9245) (31.0,58.9066) (32.0,58.8933) (33.0,58.881) };
\addlegendentry{B-tree LS, $B=256$};
\addplot coordinates { (27.0,61.7664) (28.0,61.7721) (29.0,61.776) (30.0,61.7754) (31.0,61.776) (32.0,61.7769) (33.0,61.7779) };
\addlegendentry{B-tree LS, $B=64$};
\addplot coordinates { (27.0,94.7235) (28.0,94.7228) (29.0,94.7218) (30.0,94.7224) (31.0,94.7216) (32.0,94.7219) (33.0,94.7221) };
\addlegendentry{B-tree LS, $B=8$};
\addplot coordinates { (27.0,59.7082) (28.0,59.7051) (29.0,59.7061) (30.0,59.7044) (31.0,59.7011) (32.0,59.7015) (33.0,59.7019) };
\addlegendentry{B-tree BS, $B=128$};
\addplot coordinates { (27.0,78.5999) (28.0,78.6022) (29.0,78.6019) (30.0,78.6003) (31.0,78.6003) (32.0,78.5985) (33.0,78.5991) };
\addlegendentry{B-tree BS, $B=16$};
\addplot coordinates { (27.0,58.9917) (28.0,58.9611) (29.0,58.9401) (30.0,58.9245) (31.0,58.9066) (32.0,58.8933) (33.0,58.881) };
\addlegendentry{B-tree BS, $B=256$};
\addplot coordinates { (27.0,61.7664) (28.0,61.7721) (29.0,61.776) (30.0,61.7754) (31.0,61.776) (32.0,61.7769) (33.0,61.7779) };
\addlegendentry{B-tree BS, $B=64$};
\addplot coordinates { (27.0,94.7235) (28.0,94.7228) (29.0,94.7218) (30.0,94.7224) (31.0,94.7216) (32.0,94.7219) (33.0,94.7221) };
\addlegendentry{B-tree BS, $B=8$};
\addplot coordinates { (27.0,125.64) (28.0,125.643) (29.0,125.643) (30.0,125.641) (31.0,125.641) (32.0,125.638) (33.0,125.639) };
\addlegendentry{Fusion Tree SIMD, $k=16$};
\addplot coordinates { (27.0,126.654) (28.0,126.653) (29.0,126.652) (30.0,126.653) (31.0,126.652) (32.0,126.652) (33.0,126.652) };
\addlegendentry{Fusion Tree SIMD, $k=8$};
\addplot coordinates { (27.0,125.64) (28.0,125.643) (29.0,125.643) (30.0,125.641) (31.0,125.641) (32.0,125.638) (33.0,125.639) };
\addlegendentry{Fusion Tree LS, $k=16$};
\addplot coordinates { (27.0,126.654) (28.0,126.653) (29.0,126.652) (30.0,126.653) (31.0,126.652) (32.0,126.652) (33.0,126.652) };
\addlegendentry{Fusion Tree LS, $k=8$};
\legend{}
\end{axis}
\end{tikzpicture}
\centering\ref{leg:fusion:u40}
\caption{Throughputs for the insert, predecessor and delete operations, as well as memory usage of the fusion and B-trees for 32-bit (top) and 40-bit keys (bottom). Best viewed in colour.}
\label{fig:fusion:u40}
\end{figure}
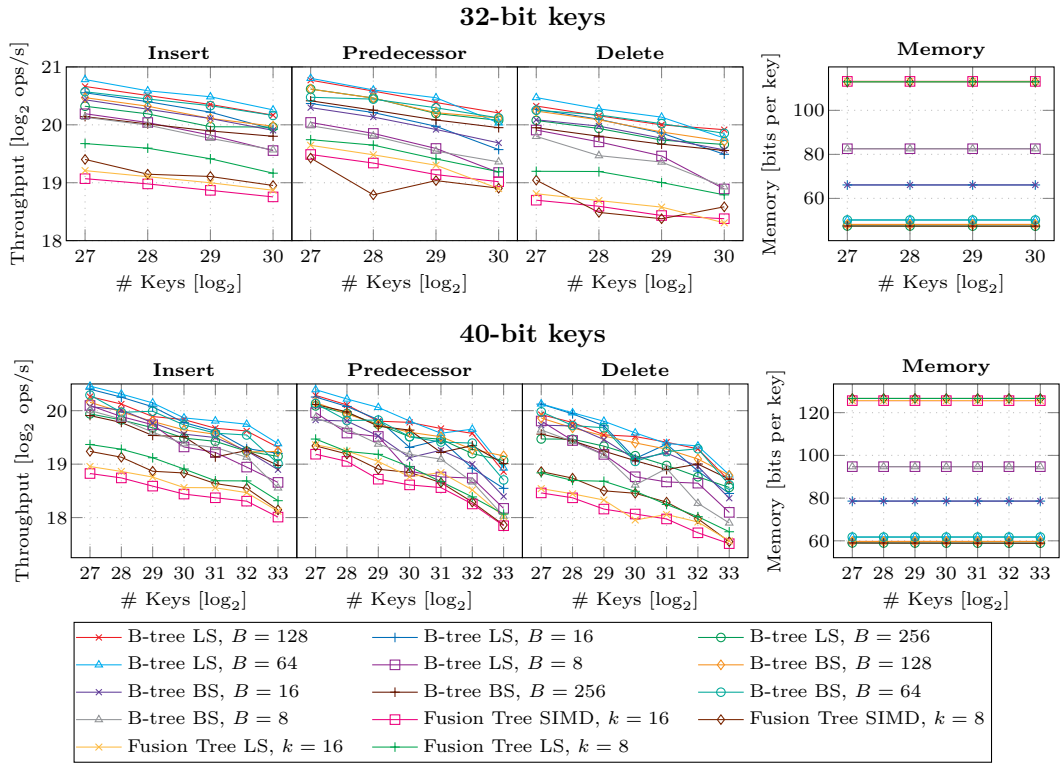

\section{Additional Results}
\label{appendix:results}
We give experimental results in addition to those presented in Sections~\ref{sec:sampling} through \ref{sec:fusion}, which have been omitted there for the sake of clarity as they lead to largely the same conclusions.
\autoref{fig:yfast:u40} shows results for our y-fast tries (\autoref{sec:yfast}) for 32-bit and 40-bit universes, respectively,
\autoref{fig:fusion:u40} for our fusion and B-trees (\autoref{sec:fusion}).

\section{Choosing Parameters For Universe Sampling}
\label{appendix:sampling-params}
The key question for preparing the experiments for our sampling data structure (\autoref{sec:sampling}) is how to configure the bucket size $b$, which is a direct trade-off parameter for the performance of the top level versus that of the bucket level:
at the top level, we have worst-case costs of $\Oh{u/b}$ for updates or queries depending on the chosen data structure.
At the bucket level, we have costs of $\Oh{b}$ for queries.
To that end, we want to pick $b$ large enough so that the top level does not end up with too many entries, and pick it small enough so that operations on the bucket level do not take too much time.

We first do some considerations on the top level.
When we assume a uniform distribution of keys inserted into the data structure, we observe that the number of insertions required until every bucket is active is distributed geometrically.
What follows is that long scans for active buckets on the top level occur more and more rarely as more keys are inserted into the data structure.
The same conclusion can be drawn when inserted keys are skewed towards a range within $U$, because then it occurs rarely that active buckets far away from others need to be accessed.
Therefore, assuming that a large enough number of keys is going to be inserted so that long top-level scans occur rarely, we focus on bucket-level performance.

On the bucket level, we have to consider our three strategies for maintaining the keys.
First, when using an unsorted list, smaller buckets clearly result in faster query times.
In preliminary experiments, the performance declined only marginally up to a bucket size of $b_\text{list} := 2^{10}$ keys, whereas buckets any larger caused a significant drop.
When using a bit vector, we benefit from scanning through bits packed into words, resulting in large buckets of $b_\text{bv} := 2^{24}$ keys still performing very well.
Here, choosing larger buckets caused insertions and deletions to become slower.
Because these operations simply mean setting or clearing bits, this effect can be explained by a higher number of cache misses.
In the hybrid case, the initial notion is that we want the unsorted list to never consume more memory than the bit vector and thus switch to when a bucket exceeds $b/\lg b$ keys.
As we desire to maintain the sweet spot threshold of $\theta_{\max} := b_\text{list} = 2^{10}$ for unsorted lists, we seek $b_\text{hybrid}$ such that $b_\text{hybrid} / \lg b_\text{hybrid} > 2^{10}$.
This is the case for $b_\text{hybrid} \geq 2^{14}$, such that the bucket size of $b_\text{bv} = 2^{24}$ is again an option.
However, consider the case where the bucket is switched from an unsorted list representation to a bit vector: we want to avoid a sudden explosion of memory occupied---and potentially wasted---by a bucket.
Because a \emph{perfect} choice of $b_\text{hybrid}$ cannot be done without any prior knowledge about the input, we explore different configurations.

We add here that we also tried \emph{sorted} lists for maintaining the keys in the buckets, enabling binary search to speed up queries.
However, the performance of updates greatly suffered and for smaller buckets, the query speedup compared to linear scans became marginal.
Sorted lists are therefore not considered in our experiments.

\section{Elaboration On Dynamic Fusion Nodes}
\label{appendix:fusion}

We expand on the description of dynamic fusion nodes from \autoref{sec:fusion}.
Let $\hat{x}^{\bitdontcare}$ indicate a compressed key that may contain don't cares.
The $k \times k$ matrix $\hat{S}^{\bitdontcare}$ over the alphabet $\{\bitzero, \bitone, \bitdontcare\}$ is represented by two $k \times k$ binary matrices \wbranch{} and \wfree{} defined as follows:
\begin{align*}
\wfree_{ij}   &= \begin{cases} 0 & \text{, if } \hat{S}^{\bitdontcare}_{ij} \neq \bitdontcare \\ 1 & \text{, if } \hat{S}^{\bitdontcare}_{ij} = \bitdontcare \end{cases}
&
\wbranch_{ij} &= \begin{cases} \hat{S}^{\bitdontcare}_{ij} & \text{, if } \wfree_{ij} = 0 \\ 0 & \text{, if } \wfree_{ij} = 1 \end{cases}
\end{align*}
Intuitively, \wfree{} identifies the don't care bits in $\hat{S}^{\bitdontcare}$, and $\wbranch_{ij}$ is either equal to $\hat{S}^{\bitdontcare}_{ij}$ if a bit is used for branching, or zero if it is a don't care bit.
The concatenation of the bits on the $i$-th \emph{row} of $\hat{S}^{\bitdontcare}$ represent the compressed (with don't cares) $i$-th key contained in $S$.

We can find the predecessor of some key $x \in U$ by determining its rank $i < k$ among the keys in $S$.
In the following, we reduce this rank query to compressed keys.
To that end, we assume that $\hat{S}^{\bitdontcare}$ is maintained such that the rows are in ascending order.
If this is not the case, we can afford to maintain an index as described in \cite{DBLP:journals/iandc/AjtaiFK84} without worsening the asymptotically constant query and update times.
We now seek the number $i'$ of the row in $\hat{S}^{\bitdontcare}$ that corresponds to the rank of $\hat{x}$ among the compressed keys.
We say that $\hat{x} \in \{0,1\}^k$ \emph{matches} a compressed key $\hat{y}^{\bitdontcare} \in \{0,1,\bitdontcare\}^k$ with don't cares if all non-don't care bits in $\hat{y}^{\bitdontcare}$ are equal to the corresponding bits in $\hat{x}$.
Formally, this is the case if $\forall j < k:\hat{y}^{\bitdontcare}\bitaccess{j} = \bitdontcare \lor \hat{x}\bitaccess{j} = \hat{y}^{\bitdontcare}\bitaccess{j}$.
We define the operation $\qmatch(x)$ that, simultaneously for all $j < k$, tests whether $\hat{x}$ matches the compressed key encoded in the $j$-th row of $\hat{S}^{\bitdontcare}$ and reports the smallest $j$ where this is not the case.
\Patrascu{} and Thorup \cite{DBLP:conf/focs/PatrascuT14} show how to perform this operation in constant time by
\begin{enumerate}[(1)]
\item computing $\hat{S}^{\hat{x}}$ by replacing the don't care bits in $\hat{S}^{\bitdontcare}$ by the corresponding bits of the $k \times k$ bit matrix $\hat{x}^k$ that contains $k$ copies of $\hat{x}$ and
\item performing a parallel row-wise greater-than comparison of $\hat{S}^{\hat{x}}$ against $\hat{x}^k$.
\end{enumerate}

We then have $i' = \qmatch(x)$.
If $x \in S$, then $i'$ is also the rank of $x$ within $S$ as shown in \cite{DBLP:journals/jcss/FredmanW93}.
Therefore, if $x = S[i']$, we already found the predecessor of $x$ after one match operation.
However, if $x \neq S[i']$, it is $x \notin S$.
In the trie, consider the ancestor of the leaf of $x$---if $x$ were contained in $S$---at level $j = \msb(x \oplus S[i'])$.
At this node, we branched off in a direction that does not necessarily lead us to the predecessor of $x$.
To see examples of this, refer to Examples~\ref{example:fusion-node-25} and \ref{example:fusion-node-4} below.
To find the actual rank $i$ of $x$ within $S$ and thus the predecessor $S[i]$ of $x$, we simulate the necessary trie navigation by performing another match operation.
Consider the case where $x < S[i']$.
In the trie, we navigate up to the lowest ancestor $v$ that has two children and take the path to the rightmost leaf in the left subtrie of $v$.
An equivalent approach is to take the path to the leftmost leaf in the right subtrie of $v$, and subtract one from that leaf's rank.
The latter can be simulated by computing $i = \qmatch(x \land 1^{w-j}0^{j})-1$.
In the case that $x > S[i']$, symmetrically, we simulate navigation to the rightmost leaf in the left subtrie of $v$ by computing $i = \qmatch(x \lor 0^{w-j}1^{j})$.

\Patrascu{} and Thorup further show how to perform all the necessary manipulations of $\hat{S}^{\bitdontcare}$ in constant time in order to insert keys into the data structure.
The intuition is always that $\hat{S}^{\bitdontcare}$ is stored in two words and all required word operations can be done in constant time.

\begin{example}
\label{example:fusion-node-25}
We consider a predecessor search for $x=25$ in the set $S$ from \autoref{fig:dynamic-fusion-node} with $w=5$.
The binary representation of $x$ is $\texttt{11001}$, which we compress to $\hat{x} = \texttt{111}$.
We compute $i' = \qmatch(x) = 4$, corresponding to $S[i']=27$.
This cannot be the predecessor of $x$ because $x < S[i']$.
The position at which we branched off in the wrong direction is $j = \msb(x \oplus y) = 2$, at the node two levels above the leaf labeled by 27 in \autoref{fig:trie}:
a path leading to $x=25$ would branch off to the left, whereas we branched off to the right.
We simulate the necessary trie navigation by computing $i = \qmatch(x \land 1^{w-j}0^{j})-1 = \qmatch(\texttt{11000})-1 = 3$.
Now, $S[i]=12$ is the correct predecessor of $x$.
\end{example}

\begin{example}
\label{example:fusion-node-4}
We consider a predecessor search for $x=4$ in the set $S$ from \autoref{fig:dynamic-fusion-node} with $w=5$.
The binary representation of $x$ is $\texttt{00100}$, which we compress to $\hat{x} = \texttt{000}$.
We compute $i' = \qmatch(x) = 1$, corresponding to $S[i']=2$.
Since $x > S[i']$, we have the opposite case as in \autoref{example:fusion-node-25}.
The position at which we branched off in the wrong direction is $j = \msb(x \oplus y) = 2$, at the node two levels above the leaf labeled by 2 in \autoref{fig:trie}.
We compute $i = \qmatch(x \lor 0^{w-j}1^{j}) = \qmatch(\texttt{00111}) = 2$, and $S[i]=3$ is the predecessor of $x$.
\end{example}

\paragraph*{Deleting keys.}
\Patrascu{} and Thorup thoroughly describe the process of inserting a key into a dynamic fusion node in constant time \cite{DBLP:conf/focs/PatrascuT14}.
They further claim that to delete a key, ``we just have to invert the [\dots] process''.
However, some details require special attention, which is why we sketch the constant-time deletion of a key here.
To that end, we use the same bag of tricks to perform the necessary $k \times k$ matrix manipulations in constant time using word operations and refer to \cite{DBLP:conf/focs/PatrascuT14} for the ideas.

Consider deleting a key $x \in U$ from our set $S$ containing at most $k$ keys from $U$.
Recall that $S$ is stored in a $k \times k$ matrix $\hat{S}^{\bitdontcare}$ of bits and don't cares represented by two words \wbranch{} and \wfree{}, and that we maintain the $k$ distinguishing positions by setting the corresponding bits in a mask $M$ of $w$ bits.
We first compute the rank $i = \qmatch(x)$ of $x$ within $S$ in constant time.
To verify that $x \in S$, we compare $x$ against $S[i]$.
If $x \notin S$, we abort the deletion.
Otherwise, we require the position $j$ of the least significant \emph{distinguishing} bit at which $x$ branches off in the trie.
This position may no longer be distinguishing after the deletion of $x$ and the, the corresponding bit must be removed from all remaining compressed keys in $\hat{S}^{\bitdontcare}$ to retain optimal compression.
If $j$ remains a distinguishing position, we need to replace the corresponding bits in all keys in the subtrie beneath node $v$ by don't cares, where $v$ is the ancestor of the leaf corresponding to $x$ on level $j$.
This is because due to the deletion of $x$, $v$ loses a child and is no longer a branching node.
We will not consider any distinguishing positions of significance higher than $j$, because for those, by construction, there must be at least one other key in $S$ that branches off the corresponding trie node.
The deletion of $x$ works as follows:

\begin{enumerate}
\item Find the position $j$ of the least significant distinguishing bit at which $x$ branches off in the trie.
      This corresponds to the position of the least significant non-don't care bit in $\hat{x}^{\bitdontcare}$.
      Compute $h$ by adding one to the number of trailing don't cares in $\hat{x}^{\bitdontcare}$, which equals the number of trailing ones in the $i$-th row of \wfree{}.
      Then, $j = \qselect_1(M, h)$.
\item Remove the $i$-th row, which contains $\hat{x}^{\bitdontcare}$, from $\hat{S}^{\bitdontcare}$.
\item Test if the deletion of $x$ results in $j$ no longer being a distinguishing position.
      This is the case if all non-don't care bits in the $h$-th column have the same value, indicating that there are no more branches at any node in the trie on level $j$.
      This can be done in constant time using a sequence of word operations; we refer to our code for details.
\item If that is the case, remove the $h$-th column from $\hat{S}^{\bitdontcare}$ and clear the $j$-th bit in $M$.
\item Otherwise, if $j$ remains distinguishing, find the range $i_0$ to $i_1$ of keys in the subtree beneath $j$.
      This range must contain at least one key, because otherwise column $h$ in row $i$ would have been a don't care.
      For all compressed keys in the range, column $h$ must be updated to a a don't care.
\end{enumerate}
Compared to \cite{DBLP:conf/focs/PatrascuT14}, we introduced two additional operations on words: counting trailing ones and a binary select operation.
In \autoref{sec:prelim}, we already mentioned briefly how these can be performed in constant time both in theory and practice.

\begin{example}
We consider the deletion of key $x=12$ in $\autoref{fig:dynamic-fusion-node}$.
It has rank $i = 3$ and occupies the third row in the matrix.
We follow the steps of our sketched algorithm:
\begin{enumerate}
\item Observe that $\hat{x}^{\bitdontcare} = \bitzero\bitone\bitdontcare$ has one trailing don't care, so we have $h=2$.
The position of the least significant distinguishing bit at which $x$ branches off is thus $j = \qselect_1(M, h) = 2$.
This corresponds to the second level in the trie shown in \autoref{fig:trie}.
\item We remove the third row from $\hat{S}^{\bitdontcare}$, conceptually removing the leaf for $x=12$ in the trie.
\item Observe how all non-don't care bits in column $h=2$ of the matrix now have the same value 0.
This corresponds to the fact that in the trie, on level $h=2$, there are no longer any branches, which means that position $j$ is no longer distinguishing.
\item Because $j$ is no longer distinguishing, we remove the second column from the matrix completely and clear the corresponding bit in $M$.
\end{enumerate}

The mask indicating distinguishing bits is now $\texttt{10001}$, and the matrix now consists of the compressed keys $\texttt{00}$ (for key 2), $\texttt{01}$ (for key 3) and $\bitone\bitdontcare$ (for key 27).
\end{example}

\end{document}